\documentclass[12pt,letterpaper]{JHEP3}

\usepackage{amscd,amsmath,amsfonts,amssymb,amsthm,mathrsfs,bbm,bm}
\usepackage{pdfsync}
\usepackage{graphicx}
\usepackage[latin1]{inputenc}
\usepackage[all]{xy}
\usepackage{stmaryrd}
\usepackage{rotating}
\usepackage{undertilde}
\usepackage{color}
\usepackage[dvipsnames]{xcolor}
\let\normalcolor\relax

\newcommand{\bit}{\begin{itemize}}
\newcommand{\eit}{\end{itemize}}
\newcommand{\bd}{\begin{description}}
\newcommand{\ed}{\end{description}}

\newcommand{\bc}{\begin{center}}
\newcommand{\ec}{\end{center}}

\def\hf{\frac{1}{2}}

\def\p{\partial}
\newcommand{\nn}{\nonumber}
\newcommand{\de}{\mathrm{d}}
\newcommand{\I}{\mathrm{i}}
\def\Re{\,\mathrm{Re}\,}
\def\Im{\,\mathrm{Im}\,}

\newcommand{\C}{{\mathbb C}}

\newcommand{\cL}{{\mathcal L}}
\newcommand{\cH}{{\mathcal H}}

\newcommand{\cE}{{\mathcal E}}

\newcommand{\cG}{{\mathcal G}}

\newcommand{\scr}{\scriptscriptstyle\rm}

\renewcommand{\sl}{{\mathfrak{sl}}}

\newcommand{\be}{\begin{equation}}
\newcommand{\ee}{\end{equation}}
\newcommand{\bea}{\begin{eqnarray}}
\newcommand{\eea}{\end{eqnarray}}
\newcommand{\bs}{\begin{subequations}}
\newcommand{\es}{\end{subequations}}

\newcommand{\w}{\wedge}

\newcommand{\f}{\frac}
\newcommand{\tl}{\tilde}
\def\p{\partial}
\newcommand{\Id}{\mathbbm{1}}

\newcommand{\na}{\nabla}

\def\({\left(}
\def\){\right)}
\def\[{\left[}
\def\]{\right]}

\def\a{\alpha}
\def\b{\beta}
\def\g{\gamma}
\def\d{\delta}
\def\eps{\epsilon}
\def\veps{\varepsilon}

\def\l{\lambda}
\def\m{\mu}
\def\n{\nu}

\def\r{\rho}
\def\vr{\varrho}
\def\s{\sigma}

\def\vp{\varphi}

\def\om{\omega}

\def\Si{\Sigma}

\def\L{\Lambda}
\def\Om{\Omega}

   \def\CA {{\cal A}}
   \def\CB {{\cal B}}
   
   \def\CD {{\cal D}}
   \def\CE {{\cal E}}
   
   \def\CG {{\cal G}}
   \def\CH {{\cal H}}
   \def\CI {{\cal I}}
   
   \def\CK {{\cal K}}
   \def\CL {{\cal L}}

   \def\CP {{\cal P}}
   \def\CQ {{\cal Q}}
   \def\CR {{\cal R}}

   \def\CV {{\cal V}}

\def\crH{\mathscr{H}}

\newcommand{\tSig}{\lefteqn{\smash{\mathop{\vphantom{<}}\limits^{\;\sim}}}\Sigma}
\newcommand{\tX}{\lefteqn{\smash{\mathop{\vphantom{<}}\limits^{\;\sim}}}X}
\newcommand{\tE}{\lefteqn{\smash{\mathop{\vphantom{<}}\limits^{\;\sim}}}E}

\newcommand{\tP}{\lefteqn{\smash{\mathop{\vphantom{<}}\limits^{\;\sim}}}P}
\newcommand{\tQ}{\lefteqn{\smash{\mathop{\vphantom{<}}\limits^{\;\sim}}}Q}
\newcommand{\tcP}{\lefteqn{\smash{\mathop{\vphantom{<}}\limits^{\;\sim}}}\CP}
\newcommand{\tcQ}{\lefteqn{\smash{\mathop{\vphantom{<}}\limits^{\;\sim}}}\CQ}

\newcommand{\Sigt}{\lefteqn{\smash{\mathop{\vphantom{\Bigl(}}\limits_{\sim}\atop \ }}\Sigma}
\newcommand{\Et}{\lefteqn{\smash{\mathop{\vphantom{\Bigl(}}\limits_{\sim}\atop \ }}E}

\newcommand{\Nt}{\lefteqn{\smash{\mathop{\vphantom{\bigl(}}\limits_{\sim}\atop \ }}N}
\newcommand{\Mt}{\lefteqn{\smash{\mathop{\vphantom{\Bigl(}}\limits_{\,\sim}\atop \ }}M}

\newcommand{\teps}{\tilde{\eps}}

\newcommand{\epst}{\lefteqn{\smash{\mathop{\vphantom{\Bigl(}}\limits_{\!\scriptstyle{\sim}}\atop \ }}\eps}

\newcommand{\htSig}{\lefteqn{\smash{\mathop{\vphantom{<}}\limits^{\;\hat\sim}}}\Sigma}
\newcommand{\htP}{\lefteqn{\smash{\mathop{\vphantom{<}}\limits^{\;\hat\sim}}}P}

\def\hCH{\hat\CH}

\def\hE{\hat E}

\def\hphi{\hat \phi}
\def\hlam{\hat \lambda}

\def\lapse{\Nt}
\def\shift{N}

\def\cN{N}
\def\Htot{H_{\rm tot}}

\def\xc{x}
\def\yc{y}
\def\ab{\beta_\gamma}
\def\abb{\alpha_\gamma}
\def\aE{\alpha}
\def\aP{\beta}

\def\opP{\mathscr{P}}

\def\lamp{\lambda_+}
\def\lamm{\lambda_-}
\def\lampm{\lambda_\pm}

\newcommand{\og}[1]{\overset{\scriptscriptstyle g}{#1}{}}
\newcommand{\ov}[2]{\overset{\scriptscriptstyle {#2}}{#1}{}}

\def\eq{e^{(1)}}
\def\eqi#1{e^{(1)#1}}
\def\mq{m^{(1)}}

\def\tEqi#1{\tE^{(1)#1}}
\def\piq{\pi^{(1)}}
\def\omq{\omega^{(1)}}
\def\omqi#1{\omega^{(1)#1}}
\def\Lamq{\Lambda^{(1)}}
\def\lamq{\lambda^{(1)}}

\def\zetaqi#1{\zeta^{(1)#1}}
\def\phiq{\phi^{(1)}}
\def\phiqi#1{\phi^{(1)#1}}
\def\psiq{\psi^{(1)}}

\def\CEq{\CE^{(1)}}
\def\CHq{\CH^{(1)}}
\def\CRq{\CR^{(1)}}
\def\CKq{\CK^{(1)}}
\def\CDq{\CD^{(1)}}
\def\crHq{\crH^{(1)}}
\def\Aq{A^{(1)}}
\def\Aqi#1{A^{(1)#1}}
\def\Bq{B^{(1)}}
\def\Bqi#1{B^{(1)#1}}
\def\Phiq{\Phi^{(1)}}
\def\Psiq{\Psi^{(1)}}

\title{\bf Canonical structure of minimal varying $\Lambda$ theories}

\author{
Sergei Alexandrov$^{1}$, Simone Speziale$^2$, Tom Zlosnik$^3$
\\
$^1${\it Laboratoire Charles Coulomb (L2C), Universit\'e de Montpellier,
CNRS, F-34095, Montpellier, France}\\
$^2${\it Aix Marseille Univ, Universit\'e de Toulon, CNRS, CPT, Marseille, France }\\
$^3${\it CEICO, Institute of Physics of the Czech Academy of Sciences, Na Slovance 1999/2, 182 21, Prague}
}

\date{\today}

\abstract{
Minimal varying $\Lambda$ theories are defined by an action built from the Einstein-Cartan-Holst first order action
for gravity with the cosmological constant $\Lambda$ as an independent scalar field,
and supplemented by the Euler and Pontryagin densities multiplied by $1/\Lambda$.
We identify the canonical structure of these theories which turn out to represent an example of irregular systems.
We find five degrees of freedom on generic backgrounds and for generic values of parameters,
whereas if the parameters satisfy a certain condition (which includes the most commonly considered Euler case)
only three degrees of freedom remain. On de Sitter-like backgrounds the canonical structure changes, and
due to an emergent conformal symmetry one degree of freedom drops from the spectrum.
We also analyze the self-dual case with an holomorphic action depending only on the self-dual part of the connection.
In this case we find two (complex) degrees of freedom, and further discuss the Kodama state,
the restriction to de Sitter background and the effect of reality conditions.
}

\begin{document}

\section{Introduction}

Formulating gravity in first-order tetrad variables has the elegant consequence that the Lagrangian
can be written entirely in the language of differential forms.
There are only six possible terms that satisfy this property,
in contrast with the infinite ones that can be considered in the metric formalism.
These are the Einstein-Cartan term including a cosmological constant $\L$,
the dual Holst term, and the Pontryagin, Euler and Nieh-Yan invariants.
These four additional terms have no impact on the local field equations,
which are equivalent to general relativity (GR) if the tetrad is invertible.
In particular, the Bianchi identities immediately imply that  $\L$ must be a constant,
just like in metric GR (see e.g. \cite{Smolin:2009bj}).
Cosmological modelling encourages on the other hand to
investigate the possibility of a spacetime-dependency of $\L$.
This is usually achieved by treating it as an independent scalar field with its own kinetic term,
see e.g. \cite{Peebles:1987ek,Wetterich:1994bg,Overduin:1998zv,Zlatev:1998tr}.\footnote{This possibility is distinct
from promoting constants such as $\Lambda$ and Newton's constant $G$ to dynamical fields
and recovering their constancy from new equations of motion \cite{Henneaux:1989zc,Jirousek:2020vhy}.}
The aforementioned formulation of gravity using tetrads suggests an alternative way
to make $\L$ dynamical without giving it an explicit kinetic term,
but assuming instead a linear coupling of $1/\L$ to the topological
densities \cite{Alexander:2018djy,Alexander:2019ctv}. The resulting terms in the action are no longer topological,
and lead to a theory with non-vanishing torsion. This allows in turn a spacetime-dependent $\L$
to be compatible with the Bianchi identities. In particular, one can find de Sitter-like solutions with spacetime-dependent $\L$
which can then be used for phenomenological applications.
We refer to these theories as minimal varying $\L$ theories, since a non-trivial spacetime dependence of $\L$
is dynamically possible even without an explicit kinetic term for it. They have the valuable feature of being more theoretically
constrained than theories with a proper kinetic term,
since they typically depend on fewer free couplings than the latter models.

In these theories, the scalar field $\L$ is not necessarily an independent degree of freedom because it
can be eliminated in favour of torsion via an algebraic equation.
Nonetheless, it is natural to expect the presence of new degrees of freedom with respect to GR
since the canonical structure is modified.
This expectation is supported by mini-superspace cosmological models \cite{Alexander:2019wne,Magueijo:2019vmk}.
Their study showed the existence of different sectors, or branches, with different
canonical structures, typically associated to a different number of degrees of freedom.
It also hinted at a special role that the Kodama state can play \cite{Kodama:1990sc,Alexander:2018djy,Magueijo:2020ntm},
thus reviving older ideas on a duality between the cosmological constant and a notion of Chern-Simons time \cite{Smolin:1994qb}.
Clarifying these properties requires a full canonical analysis, which is what we present in this paper.

We consider a general theory with independent coupling constants in front of
the Holst, Euler and Pontryagin terms, which we denote by $1/\gamma$, $\aE$ and $\aP$, respectively.
We do not include a further coupling of $1/\L$ to the Nieh-Yan term since it would
rule out the de Sitter solution, which is on the other hand important
for phenomenological applications of theories of this type.\footnote{We nonetheless point out
that a logarithmic coupling would preserve the de Sitter solution.}
The constant $\gamma$ is commonly known as Immirzi or Barbero-Immirzi parameter \cite{Immirzi:1996dr,Barbero:1994ap}.
The Euler-only theory initially considered in \cite{Alexander:2019ctv} and mostly discussed in the literature
appears as a special case with $1/\g=\beta=0$.

The main part of the paper is devoted to the canonical analysis of the general theory.
We first present the general Lorentz covariant expression of the 3+1 decomposed action,
but then restrict the subsequent analysis to the time gauge which fixes the internal boost transformations and is always admissible.
It has the advantage of simplifying all computations, without affecting the canonical structure.
What emerges from the analysis is that the seemingly minor modification of turning the cosmological constant into an independent
scalar field and coupling it to the topological densities, has significant consequences for the
dynamics.
For generic values of the parameters, the theory propagates 5 degrees of freedom, in contrast with the 2 of GR,
or the 3 that one may naively expect adding a new scalar field.
This result depends crucially on the non-vanishing of a  certain 3-vector, which is proportional to a combination of
parameters of the theory and to the spatial gradient of $\L$.
For varying $\L$, the vector vanishes when
\be
\left(1-\f1{\g^2}\right)\b=\f{2\a}\g\,.
\label{condparam}
\ee
For these values of the parameters, we show that the theory has only 3 degrees of freedom.
Note that the condition \eqref{condparam} implies that if the Holst term is removed,
the Pontryagin term must be removed as well, and thus includes the Euler theory.
It also includes a theory with the Pontryagin term and no Euler, but the Immirzi parameter fixed to one.

In all cases, the additional degrees of freedom appearing on top of the massless graviton can be seen to reside in torsion.
While in GR the torsion vanishes and six of the corresponding equations
appear as secondary (second class) constraints,
in this model only a part of the constraints survives. The precise number of the secondary constraints depends
on the 3-vector mentioned above, and it is less than six in all cases.

These results hold for generic points in phase space, however \emph{not}
for perturbations around the  varying-$\L$ de Sitter solutions.
It turns out that the phase space of these theories contains subspaces where the canonical structure
differs from the one at generic points, and de Sitter is one example.
Dynamical systems in which this occurs are known as {\it irregular}.
The de Sitter solutions exist only for special values of the parameters given by
\be
\alpha=-\f32,\qquad \beta=-\f3{2\g}.
\label{condparamdS}
\ee
What makes this case stand out is that it has an extra symmetry given by conformal rescalings.
It manifests in the linearized approximation around such backgrounds as an additional gauge symmetry.
Its presence changes the canonical structure, and leads to one fewer degrees
of freedom for all allowed values of the parameters.
That is, we find 4 degrees of freedom in the generic case  \eqref{condparamdS},
and 2 degrees of freedom if the condition \eqref{condparam} is further satisfied,
which implies in the end the unique case with $1/\g=\beta=0$.
This reduction of degrees of freedom confirms and generalizes what was found in mini-superspace analysis
\cite{Magueijo:2019vmk}.

Then we move to consider a version of the model corresponding to a complex self-dual theory.
This is defined taking $\gamma=-\I$, $\aP=\I\aE$, and it is described by a one-parameter family
of holomorphic Lagrangians depending only on
the self-dual projections of the fields.
Such a theory is a generalization of self-dual gravity in Ashtekar variables \cite{Ashtekar:1986yd},
and is closely related to a set of modified gravity models
extensively analyzed in \cite{Krasnov:2009iy,Ishibashi:2009wj,Herfray:2015rja,Krasnov:2020zfi} and references therein.
The canonical analysis we established for the theory with real action cannot be applied directly to this case
because the symplectic structure is singular in the limit $\g \rightarrow \pm\I$.
One needs to perform the analysis {\it ex-novo}.
A nice feature of the self-dual theory is that upon integration by parts, the momentum conjugate to $\Lambda$ can be identified
with the Chern-Simons (CS) Lagrangian, prompting the considerations of \cite{Alexander:2018djy,Alexander:2019ctv}.
Therefore, we find it instructive to perform the canonical analysis
of the self-dual theory in this alternative canonical frame obtained by integration by parts,
instead of repeating the previous calculations.
The resulting canonical structure turns out to be quite different, perhaps
unsurprisingly because of the aforementioned singular limit.
In particular, we find a single secondary constraint, of a form completely unrelated to
the constraints of the theory with real action, and 2 degrees of freedom, for any value of the free parameter $\a$.
This counting is based on treating all variables as holomorphic, so these are two
\emph{complex} degrees of freedom.

It is known \cite{Alexander:2018djy} that provided we fix $\a=-3/2$,
the self-dual theory also admits the varying-$\L$ de Sitter solution
which, as for real parameters, requires special attention.
Again one has an additional conformal symmetry visible in the linearized approximation, but in contrast to
the real case, it does not lead to a reduction of degrees of freedom. Their number remains equal to 2.
This happens because, while one of the second class constraints is converted into first class, another one drops out.
It is actually this change in the canonical structure that ensures that a proper generalization of the Kodama state
\cite{Kodama:1990sc,Alexander:2018djy}
provides a formal solution to all quantum constraints.

Finally, we discuss the reality conditions to be imposed on the self-dual theory to make it physical in Lorentzian signature.
This is known to be a thorny issue and there is in fact no guarantee
that the holomorphic canonical structure has a consistent real section.
For the original Ashtekar formulation of gravity \cite{Ashtekar:1987gu}, it is a remarkable result
that reality conditions can be consistently imposed on a holomorphic action, but
no analogous result exists for more general theories. This problem was recently analysed in
\cite{Krasnov:2020zfi}, where it was pointed out that the stability of the reality conditions should be analysed
using a {\it holomorphic} Hamiltonian
describing dynamics of the self-dual theory and {\it not} including any reality conditions.
This approach reproduces the results of \cite{Ashtekar:1987gu}, but also exposes a problem in
the dynamics of the modified theories \cite{Krasnov:2009iy,Ishibashi:2009wj,Herfray:2015rja,Krasnov:2020zfi},
leading to effectively zero degrees of freedom.
We follow this approach for the self-dual modified theory of gravity considered here and
show that the reality conditions cannot be stabilized, in the sense that at each step
of the stabilization procedure one generates new reality conditions.
Thus, their imposition appears to be inconsistent with the holomorphic dynamics.

The structure of the paper is the following. In the next section we briefly review the model,
its equations of motion and provide a detailed analysis of the varying-$\L$ de Sitter solutions.
In section \ref{sec-canan} we perform the canonical analysis.
First, we derive the 3+1 decomposition of the action, then impose the time gauge
and study the stability conditions of the primary constraints.
The analysis splits into two cases depending on
whether the parameters of the model satisfy or not the condition \eqref{condparam}.
These cases are presented in subsections \ref{case-spec} and \ref{case-gen}, respectively.
The last subsection \ref{subsec-deSitter} is devoted to the canonical analysis around the de Sitter background.
In section \ref{SecSD} we investigate the self-dual theory, the associated de Sitter solution,
and discuss the fate of the Kodama state,
as well as the effect of imposition of reality conditions.
Finally, we conclude with a discussion in section \ref{sec-concl}.
A few appendices contain details of our calculations and the constraint algebra.

Our conventions are such that the internal space indices $I,J=0,\dots,3$ are raised and lowered by means of the flat Minkowski metric
$\eta_{IJ}=\mbox{diag}(-,+,+,+)$. The Levi-Civita symbol with flat indices is normalized as $\eps_{0123}=1$, and
similarly the antisymmetric tensor density with spacetime indices is taken to satisfy
$\epst_{0abc}=\epst_{abc}$, where $a,b,\dots$ label spatial directions and
the tilde above or under tensors defines densities of weight +1 and $-1$ respectively. Spacetime indices are denoted with greek letters.
The symmetrization and anti-symmetrization of indices are denoted by $(\cdot\, \cdot)$ and $[\cdot\,\cdot]$, respectively,
and include the factors of 1/2.

\section{The model}
\label{sec-action}

The theory we consider has the following action depending on the tetrad one-form $e^I$,
the connection one-form $\omega^{IJ}$ and the scalar field $\Lambda$,
\begin{align}\label{L1L2}
S(e,\om,\L) = S_1(e,\om,\L) + S_2(e,\om,\L),
\end{align}
where
\begin{subequations}
\bea
S_1(e,\om,\L) &=&\hf\int \( \hf\,{\eps^{IJ}}_{KL} + \frac{1}{\g}\,\d^{IJ}_{KL} \) e_I\w e_J \w \(F^{KL}(\omega) - \f\L6 ~ e^K\w e^L\),
\label{defL1}
\\
S_2(e,\om,\L) &=&\hf \int \( \frac{\aE}{2}\,{\eps^{IJ}}_{KL}  + {\aP}\,\d^{IJ}_{KL} \) \frac{1}{\L}\, F_{IJ}(\omega)\w F^{KL}(\omega),
\label{defL2}
\eea
\label{twoactions}
\end{subequations}
and $F^{IJ}(\omega)$ is the curvature of the connection.
We take units $8\pi G=1$, and $\g, \aE, \aP$ are three dimensionless coupling constants.
For $\L$ constant, $S_1$ is the Einstein-Cartan-Holst action for GR in the first order formulation, and $S_2$
contains the dimension-4 terms corresponding respectively to the Euler and Pontryagin topological invariants.
The modification to GR comes from making the cosmological constant an independent scalar field.
After this modification, the second part of the action \eqref{defL2}
is not topological anymore and contributes to the equations of motion.
In particular, it gives rise to terms involving derivatives of $\Lambda$ in spite of
it does not have its own kinetic term (a feature that inspired the name of ``minimal varying $\L$" theories).
Among this class of theories, the most commonly studied is the `Euler theory',
which corresponds to adding only the Euler invariant, namely $1/\g=\b=0$.
Speaking of coupling constants, comparison of the two actions \eqref{defL1}
and \eqref{defL2} suggests that it might be convenient to parametrize the coupling of the Pontryagin term using
\be
\g':=\f\a\b.
\ee
Then both action terms can be written in terms of the operator
\be
\opP^{IJ}_{KL}(\g)= \f12\,{\eps^{IJ}}_{KL} +\f1\g\,\d^{IJ}_{KL}.
\ee
In the rest of this paper we will alternatively use the parametrization of
the couplings in terms of $\b$ or $\g'$, whichever is more convenient.

The field equations obtained varying respectively $e, \om$ and $\L$ read
\begin{subequations}\label{FE}
\begin{align}
\label{FE1}
& \opP_{IJKL}(\g) \, e^J\w F^{KL} -\f\L6\, \eps_{IJKL}e^J\w e^K\w e^L = 0,\\
\label{FE2}
& \opP_{IJKL}(\g) \, T^K \w e^L+ \a\, \opP_{IJKL}(\g') \, \de\L^{-1}\w F^{KL} = 0, \\
\label{FE3}
& \L^2\eps + \frac{\a}{2}\, \opP_{IJKL}(\g') \, F^{IJ}\w F^{KL} = 0,
\end{align}\end{subequations}
where
$T^I:= D^{(\omega)}e^I=\de e^I+{\omega^I}_J\wedge e^J$
is the torsion and
$\eps:=\f1{4!}\eps_{IJKL}e^I\w e^J\w e^K\w e^L$
is the volume 4-form.
The last equation allows us to express the scalar field $\Lambda$ through the curvature and thereby to exclude it from the system.
However, this introduces non-algebraic expressions which make the analysis more complicated.
Therefore, we will not pursue this route.

The field equations show the central role that torsion plays in this theory.
For if we set $T^I=0$, it is immediate to see that the only solution occurs for $\a=-3/2$
and is the standard de Sitter metric with constant $\L$.
Therefore these theories are very different from their metric counterparts in which one
couples a scalar field to the Gauss-Bonnet term, see e.g. \cite{Kobayashi:2011nu,Silva:2017uqg,Anson:2019uto}
where a variable cosmological constant is allowed without torsion.

\subsection{De Sitter solution and conformal symmetry}
\label{deSitter-sol}

An important property of some of these Lagrangians is that they admit
(anti) de Sitter
with spacetime-dependent $\L$ as an exact solution.
This solution satisfies
\be
F^{IJ}=\frac{\Lambda}{3}\, e^I\w e^J,
\label{deSit}
\ee
and it is valid only for specific parameters
\be
\a=-3/2,
\qquad
\gamma'=\gamma.
\label{deSit-param}
\ee
The existence of the de Sitter solution  was shown already in \cite{Alexander:2019ctv} for the Euler theory,
and we extend it here to the case including the Holst and Pontryagin terms,
which requires them to have the same coupling constant.
To prove this result, we take \eqref{deSit} as an ansatz, and plug it in the field equations.
The first equation \eqref{FE1} is manifestly solved for any $\g$, whereas \eqref{FE3}
is solved for any $\g'$ but requires $\a=-3/2$. Finally  \eqref{FE2} gives
\be\label{FE2dS}
\f12\,{\eps^{IJ}}_{KL}  \left( T^K - \f{\a}3\,\de\log\L\w e^K\right)\w e^L +\left(\f{1}{\g}\, T^I -\f{\a}{3\g'}\,\de\log\L\w e^I\right)\w e^J=0.
\ee
Before solving this equation, let us observe that applying the Bianchi identity to the ansatz \eqref{deSit} we obtain
\be
D^{(\omega)}F^{IJ} = \f13 \(\de\L\w e^I +2\L T^I\)\w e^J =0.
\ee
This implies that the torsion and $\L$ are related by
\be
T^I = -\f12\,\de \log\L\w e^I.
\label{deSit-tor}
\ee
The special form of torsion means that only its irreducible spin-1 trace part is present
(see e.g. \cite{Speziale:2018cvy,Hehl:1976kj} for a decomposition of torsion into irreducible representations),
whereas the completely antisymmetric piece is ruled out by the requirement of the de Sitter ansatz,
in agreement with the analysis of mini-superspace models \cite{Magueijo:2019vmk}.
Substituting \eqref{deSit-tor} into \eqref{FE2dS}, one observes that the first round bracket vanishes due to the
restriction $\a=-3/2$, whereas the second bracket leads to the additional requirement $\g' = \g$.

The on-shell value \eqref{deSit-tor} for the torsion allows us to express the connection in terms the tetrad and the field $\Lambda$.
This in turn determines the curvature $F^{IJ}$ in terms of the Riemann tensor and derivatives of $\L$.
Reinserting this expression for the curvature back into the ansatz
\eqref{deSit}, one arrives at the equations on the tetrad and $\Lambda$
which remain to be solved to find the solution explicitly. In Appendix~\ref{ap-deSitter} we show that these equations reduce
to the following two conditions,
\be
{C}^\m{}_{\n\r\s}(\L g) =0,
\qquad
{R}_{\m\n}(\Lambda g) = \L g_{\mu\nu},
\label{Weyl-eq}
\ee
where ${C}^\m{}_{\n\r\s}(\Lambda g)$ is the Weyl tensor of the metric $g_{\mu\nu}=\eta_{IJ} e_\mu^I e_\nu^J$ rescaled by $\Lambda$
and ${R}_{\m\n}(\Lambda g)$ is the Ricci tensor of the same metric.
This result shows that $\Lambda$ remains undetermined and can be chosen freely since it can be absorbed into a rescaling of the metric.

This observation has the following explanation.
For generic parameters, the theory \eqref{L1L2} is clearly invariant under
the usual diffeomorphism symmetry and local Lorentz transformations in the tangent space.
They exhaust local symmetries of the model. However, it was observed in \cite{Magueijo:2019vmk} that,
restricted to the sector \eqref{deSit},
the theory acquires additional conformal symmetry:
\be
\om^{IJ}\mapsto \om^{IJ},
\qquad
e^I\mapsto \Om e^I,
\qquad
\L\mapsto \Om^{-2}\L.
\label{conftr}
\ee
It is clear that this symmetry allows to absorb $\Lambda$ and set it to 1.
The conformal symmetry can be either checked at the level of equations of motion, or seen directly from the action.
Indeed, for the parameters given in \eqref{deSit-param}, the action \eqref{L1L2} can be rewritten as
\be
S=-\frac{3}{4}\int \frac{\opP_{IJKL}(\g)}{\Lambda}
\(F^{IJ}-\frac{\Lambda}{3}\, e^I\wedge e^J\)\wedge \(F^{KL}-\frac{\Lambda}{3}\, e^K\wedge e^L\).
\label{actdual}
\ee
Each of the two brackets is both invariant under the transformation \eqref{conftr}
and vanishes at the de Sitter solution so that the variation
of the action evaluated on this solution indeed vanishes.
Note that this form of the action makes manifest the duality symmetry \cite{Alexander:2018djy,Alexander:2019ctv}
\be
F^{IJ}\leftrightarrow \frac{\Lambda}{3}\, e^I\w e^J,
\label{duality}
\ee
with respect to which de Sitter is a self-dual solution.

It is instructive also to realize \eqref{conftr} as a gauge symmetry of the linearized theory.
To this end, let us expand the action \eqref{actdual} around the de Sitter background up to quadratic order.
Denoting the fluctuations of the fields by the upper index $^{(1)}$, one finds
\be
S^{(2)}=-3\int \frac{\opP(\g)}{\Lambda} \(D^{(\omega)}\omq-\frac{\Lambda}{3}\, e\wedge \eq-\frac{e^2}{6}\,\Lamq \)^2,
\label{Slinear}
\ee
where we suppressed the tangent space indices.
This linearized action is clearly invariant under
\be
\delta\eq =\veps \, e,
\qquad
\delta\Lamq=-2\veps \Lambda.
\label{conftr-q}
\ee

Being a gauge symmetry, the conformal symmetry reduces the number of perturbative degrees
of freedom propagating on the de Sitter background,
and it is reminiscent of the partial massless symmetry of massive gravity
\cite{Deser:1983mm,Deser:2001us,Deser:2004ji}. Similarly to our case, that symmetry appears only in a linearized approximation around
a ceratin class of backgrounds and disappears when quadratic fluctuations are taken into account \cite{Alexandrov:2014oda}.
Such behavior is typical for irregular systems, i.e. those whose canonical structure depends on a point in the phase space.
(See e.g. \cite{Deser:2012ci,Alexandrov:2012yv} for other examples of gravitational models with irregular canonical structure.)
Below we confirm that our model is an example of such irregular system. We perform a thorough canonical analysis,
first at the non-perturbative level on a {\it generic} background and then for the case of
the linearized theory on the de Sitter background. As we will see, the canonical structures indeed turn out be different.

\subsubsection*{On the Nieh-Yan invariant}

Before moving on, let us explain why a similar coupling to the Nieh-Yan invariant rules out the de Sitter solution.
Suppose we add to the action a term
\be
S_{\scr NY}= \frac{\g_{\scr NY}}{2}\int\frac{1}{\L}\,  \de (e_I\w T^I).
\ee
The field equations \eqref{FE} are modified adding respectively the terms
\be
\g_{\scr NY}\de\L^{-1}\w T_I,
\qquad
-\f{ \g_{\scr NY}}{2}\,\de\L^{-1}\w e_I\w e_J,
\qquad
\f12\g_{\scr NY}\de(e_I\w T^I).
\ee
The de Sitter ansatz \eqref{deSit} and the Bianchi identities still imply
that torsion must have only the trace spin-1 component as in \eqref{deSit-tor}.
This makes the first and third of the new terms vanish identically, so the corresponding field equations are still satisfied.
However in the second field equation \eqref{FE2} the condition $\g'=\g$ is now replaced by
\be\label{NYcond}
\f1{\g'}-\f1\g +\f {\g_{\scr NY}}{\L}=0.
\ee
The only possibility to solve this equation with varying $\L$ is to set $\g_{\scr NY}=0$.

From this analysis, one can also deduce that the obstruction is avoided if the Nieh-Yan invariant is coupled not to
$\L^{-1}$, but to $\log\L$. The extra terms to the first and third field equations
are still identically zero on-shell of \eqref{deSit-tor},
and the third term in \eqref{NYcond} is replaced by $-\g_{\scr NY}$ alone.
In other words, the action
\be
S_1(e,\om,\L)+S_2(e,\om,\L)+\hf\left( \f1{\g'} -\f1\g \right) \int \log\L \, \de (e_I\w T^I)
\ee
admits the varying-$\L$ de Sitter solution for $\a=-3/2$ and any values of $\g$ and $\g'$.
We will not pursue a detailed analysis of this more general action in this paper.

\section{Canonical analysis}
\label{sec-canan}

\subsection{3+1 decomposition}

The starting point of the canonical analysis is to introduce a foliation of spacetime by space-like hypersurfaces
and to parametrize the tetrad according to this foliation. A convenient parametrization analogous to
the ADM decomposition of the metric \cite{Arnowitt:1960es} is given by
\be
e^I=\(\lapse \tX^I+\shift^a e_{a}^I\)\de t + e_{a}^I\de \xc^a,
\label{dec-tetrad}
\ee
where $\tX_Ie_a^I=0$, $\lapse$ and $\shift^a$ are the usual lapse and shift fields,
and up/down tilde denotes the density weight +1/$-1$.
Note also that
\be
q_{ab}=\eta_{IJ}e_a^I e_b^J
\label{indmet}
\ee
is the metric induced on space-like hypersurfaces.\footnote{An advantage of this parametrization is that it makes it possible to treat
on equal footing also the canonical analysis on time-like \cite{Alexandrov:2005ar} and null \cite{Alexandrov:2014rta} hypersurfaces
since the signature is determined by the norm of the field $\tX^I$.
It is particularly convenient in the null case, where it allows one to avoid the use of a 2+2 foliation
and problems associated with gauge fixing.}
We denote its determinant by $q$, and $\det e=\lapse q$.

Substituting the parametrization \eqref{dec-tetrad} into the action \eqref{L1L2} and integrating by parts, one obtains
\be
S = \int \de^4 \xc\[\tP^a_{IJ}\p_0\omega^{IJ}_a
+\omega_0^{IJ} \cG_{IJ} +\shift^a \CV_a+\lapse \CH\],
\label{canonact}
\ee
where the momentum conjugated to the connection is
\be
\tP^a_{IJ}=\frac12\,\teps^{abc}\( \opP_{IJKL}(\g)\, e^K_b e^L_c+\frac{\a}{\L}\, \opP_{IJKL}(\g') F^{KL}_{bc} \),
\label{covarP}
\ee
and
\begin{subequations}
\bea
\CG_{IJ}&=&\p_a\tP^a_{IJ}+{\omega_{a,I}}^{K}\tP^a_{KJ}+{\omega_{a,J}}^{K}\tP^a_{IK},
\\
\CV_a &=&\frac12\,\teps^{bcd}\opP_{IJKL}(\g)\,e_a^I\,  e^J_b F^{KL}_{cd},
\\
\CH &=&\frac12\,\teps^{abc}\,\opP_{IJKL}(\g)\, \tX^I e^J_a \Big(F^{KL}_{bc} -\f{2\L}{3}\,e^K_b e^L_c\Big).
\eea
\label{primconstr}
\end{subequations}
Since $\omega_0^{IJ}$, $\shift^a$ and $\lapse$ appear in \eqref{canonact} only linearly,
they can be interpreted as Lagrange multipliers and are excluded from the phase space.
Each of them generates a primary constraint so that we have
\be
\CG_{IJ}\approx \CV_a\approx \CH \approx 0.
\ee
As usual, the three sets of constraints are responsible for local gauge transformations,
spatial and time diffeomorphisms, respectively.
However, they do not exhaust all primary constraints of the model.
First of all, since $\Lambda$ appears in the action without time derivatives, its conjugate momentum $\pi$
must vanish,
\be
\pi\approx 0.
\ee
Secondly, the 18 components of the momenta $\tP^a_{IJ}$ \eqref{covarP} depend only on 12 components of the tetrad $e_a^I$.
This implies that there are 6 additional primary constraints, which are known in the context of GR as simplicity constraints.
In the present case they can be written as follows
\be
\phi^{ab}=\(\tP^a_{IJ}-\tcP^a_{IJ}(\omega)\)(\opP^{-1}\star \opP^{-1})^{IJKL}\(\tP^b_{KL}-\tcP^b_{KL}(\omega)\)\approx 0,
\label{covsimpl}
\ee
where we introduced
\be
\tcP^a_{IJ}(\omega):=\frac{\a}{2\L}\,\teps^{abc}\, \opP_{IJKL}(\g') F^{KL}_{bc},
\ee
and
\be
(\opP^{-1}\star \opP^{-1})^{IJ}_{KL}=\(1+\tfrac{1}{\gamma^2}\)^{-2}
\(\tfrac12\(1-\tfrac{1}{\gamma^2}\){\eps^{IJ}}_{KL}
-\tfrac{2}{\gamma}\, \delta^{IJ}_{KL}\) .
\ee

As customary from the study of GR in tetrad variables, it is convenient
to redefine the constraint $\CV_a$ so that the new constraint generates the ordinary Lie derivative on the dynamical fields.
Accordingly, we replace it by the following linear combination of primary constraints
\be
\begin{split}
\CD_a:=&\, \CV_a+\omega_a^{IJ}\CG_{IJ}-\pi\p_a\Lambda
\\
=&\, \p_b\(\omega_a^{IJ}\tP^b_{IJ}\)-\tP^b_{IJ}\p_a\omega_b^{IJ}-\pi\p_a\Lambda.
\end{split}
\label{constrD}
\ee
In obtaining the second equality we used the fact that the contribution
$\frac{\alpha}{2\Lambda}\,\teps^{bcd}\opP_{IJKL}(\gamma') F_{ab}^{IJ} F_{cd}^{KL}$,
arising from $\CV_a$ after expressing it in terms of the canonical momentum,
actually vanishes due to the identity $\teps^{bcd} F_{ab}^{IJ} F_{cd}^{KL}=-\teps^{bcd} F_{ab}^{KL} F_{cd}^{IJ}$.
The constraint $\CD_a$ \eqref{constrD} is manifestly in the standard form of the generator of spatial diffeomorphisms.

\subsection{Time gauge}

The form of the simplicity constraints \eqref{covsimpl} is highly complicated due to their quadratic nature
and the presence of the terms involving curvature.
It is possible to simplify the analysis imposing a partial gauge fixing of the internal Lorentz symmetry.
This consists of aligning one vector of the tetrad with the normal to the hypersurface, and it is known as time gauge.
Such gauge is always admissible and cannot change the canonical structure, in particular, the counting of degrees of freedom.

To impose this gauge, we follow \cite{Alexandrov:1998cu} and  parametrize
\be
\begin{split}
\tX^I= \sqrt{q} \(1,\chi^i\),
\qquad
e_{a}^I=\,\(E_{a}^j\chi_{j},E_{a}^i\),
\end{split}
\label{parameX}
\ee
which ensures the defining constraint $\tX_Ie_a^I=0$. The variable $\chi^i$ controls the normal to hypersurfaces and,
in particular, they are space-like, null, or time-like when $\chi^2=\d^{ij}\chi_i\chi_j$ is less, equal or greater than 1, respectively.
The time gauge for the space-like case corresponds to the simplest choice
\be
\chi^i=0.
\label{timegauge}
\ee
In this gauge $\sqrt{q}=\det E_a^i$ and $q_{ab}=\d_{ij} E_a^iE_b^j$.
We will also extensively use the inverse triad $E^a_i$ and its densitized version $\tE^a_i = \sqrt{q} E^a_i$.
Finally, we introduce the standard notations for the
boost and rotation components of the connection,
\be
K_a^i=\omega^{0i}_a,
\qquad
\Gamma_a^i=\hf\, {\eps^i}_{jk}\omega_a^{jk}.
\label{compom}
\ee

With these notations, the action \eqref{canonact} in the time gauge takes the following
form\footnote{The Lagrange multipliers $n^i$ and $m^i$ are related to the time components of the connection by
$$
m^i=\omega^{0i}_0 -\cN^a \omega^{0i}_a,
\qquad
n^i=\hf\,{\eps^i}_{jk}\(\omega^{jk}_0-\cN^a \omega^{jk}_a\).
$$
}
\be
S = \int \de \xc^4  \[\tP^a_i\p_0 K^i_a+\tQ^a_i\p_0\Gamma^i_a
+m^i\CK_i+n^i \CR_i +\shift^a \CD_a+\lapse \CH\],
\ee
where the two momenta are
\be
\begin{split}
\tP^a_i =&\,
\tE^a_i+\tcP^a_i(\omega),
\qquad\quad
\tcP^a_i(\omega):=\frac{\a}{\L}\,\teps^{abc}\(F^{i}_{bc}-\tfrac{1}{\gamma'}\, F_{bc}^{0i}\),
\\
\tQ^a_i  =&\,
\frac{1}{\gamma}\, \tE^a_i+\tcQ^a_i(\omega) ,
\qquad
\tcQ^a_i(\omega):=\frac{\a}{\L}\,\teps^{abc}\(F^{0i}_{bc}+\tfrac{1}{\gamma'}\, F_{bc}^{i}\),
\end{split}
\label{momenta}
\ee
and the primary constraints are given by
\begin{subequations}
\bea
\CK_i &=&
\p_a \tP^a_i+{\eps_{ij}}^k\(K_a^j\tQ^a_k-\Gamma_a^j\tP^a_k\),
\\
\CR_i &=&
\p_a \tQ^a_i-{\eps_{ij}}^k\(K_a^j\tP^a_k+\Gamma_a^j\tQ^a_k\),
\\
\CD_a&=& \p_b(K_a^i \tP^b_i)-\tP^b_i\p_a K_b^i+\p_b(\Gamma_a^i \tQ^b_i)-\tQ^b_i\p_a\Gamma_b^i-\pi\p_a\Lambda,
\\
\CH &=&\hf \,{\eps^{ij}}_k\tE^a_i \tE^b_j \(F^{k}_{ab}-\tfrac{1}{\gamma}\, F_{ab}^{0k}\) -\L q.
\label{defCHtg}
\eea
\label{primcon-tg}
\end{subequations}
The explicit expressions for the components of the curvature $F_{ab}^{0i}$ and $F_{ab}^i:=\hf\, {\eps^i}_{jk}F_{ab}^{jk}$
in terms of \eqref{compom} can be found in \eqref{compF}.
From the form of the momenta \eqref{momenta}, it is easy to see that they satisfy 9 constraints
\be
\phi^a_i= (\tQ^a_i-\tcQ^a_i)-\frac{1}{\gamma}\,(\tP^a_i-\tcP^a_i)\approx 0,
\label{primcon}
\ee
which replace the non-linear simplicity constraints \eqref{covsimpl}.
The three additional primary constrains appear due to the gauge fixing \eqref{timegauge}.
Thus, in total one has 20 primary constraints $\CK_i$, $\CR_i$, $\CD_a$, $\CH$, $\pi$ and $\phi^a_i$.

The full algebra of primary constraints is presented in appendix \ref{ap-conalg1}.
It is important to note that to compute this algebra, the constraints should be expressed through the canonical variables.
In \eqref{primcon-tg} this is already done for all constraints except $\CH$, which is still written as a function of
the densitized triad.
There are many ways to express $\tE^a_i$ in terms of the canonical variables,
which all differ by terms proportional to the primary constraint \eqref{primcon} and hence can affect the commutation relations.
Of course, the final canonical structure is independent of this choice since it boils down to a redefinition of primary constraints.
We choose the expression which is in a sense orthogonal to \eqref{primcon}, and ensures the simplest commutators. This is given by
\be
\tE^a_i= \frac{1}{1+1/\gamma^2}\(\tP^a_i-\tcP^a_i+\frac{1}{\gamma}\(\tQ^a_i-\tcQ^a_i\)\).
\label{defE2}
\ee
It is the only choice for which the Hamiltonian constraint $\CH$ is weakly commuting with both $\CK_i$ and itself.

\subsection{Stabilization of primary constraints}

The next step is to study stability conditions of the primary constraints, i.e. to require that their Poisson brackets with
the total Hamiltonian vanish. The latter is given by a combination of all primary constraints and reads as
\be
-\Htot=\CR(n)+\CK(m)+\CD(\vec\shift)+\CH(\lapse)+\pi(\rho)+\phi(\vec \lambda),
\label{Htotnew}
\ee
where we used the notation for smeared constraints
\be
\Phi(f)=\int\de\xc^3 f(\xc)\Phi(\xc).
\ee

From the algebra presented in appendix \ref{ap-conalg1}, one can immediately conclude that
$\CR_i$ and $\CD_a$ commute with all other constraints and therefore are stable under time evolution.
The stability of the other constraints is more non-trivial.
First, the stabilization of $\CK_i$ requires
\be
\{\Htot,\CK_i\}\approx -\(1+\gamma^{-2}\){\eps_{ij}}^k\lambda_a^j\tE^a_k=0,
\ee
where the only non-vanishing contribution to the l.h.s. comes from the
commutator \eqref{com-Kphi}.
This condition fixes
3 of the 9 components of the Lagrange multiplier $\lambda_a^i$ and implies that it can be represented
as $\lambda_a^i=\Et_{aj}\lambda^{ij}$ where $\lambda^{ij}$ is a symmetric matrix.
Similarly, the stabilization of the antisymmetric projection of $\phi^a_i$ given by
${\eps_{ij}}^k\Et_a^j\phi^a_k$ leads to a condition on the Lagrange multiplier $m^i$
for the constraint $\CK_i$. This is consistent with the expectation that
the boost generators $\CK_i$ and the three components of $\phi^a_i$ corresponding to the gauge fixing \eqref{timegauge}
(i.e. additional to the simplicity constraints \eqref{covsimpl})
form a pair of second class constraints.
Using the constraint algebra \eqref{2cl-alg}, one finds that the multiplier $m^i$
is fixed in terms of $\lambda^{ij}$, $\rho$ and $\lapse$ as follows
\be
m_i= \CA_i  \rho +\CB^+_i\lapse+\frac{\ab}{q\(1+1/\gamma^{2}\)}\({\lambda_i}^j\tE^a_j-{\lambda_k}^k\tE^{a}_i\)\p_a\Lambda^{-1}.
\label{fixm}
\ee
Here $\CA_i$ and $\CB^+_i$ are functions on phase space defined in \eqref{defCAB}, and
the parameter $\ab$ is given by
\be
\ab
:=\(1-\frac{1}{\gamma^2}\)\aP-\frac{2\aE}{\gamma}.
\label{defab0}
\ee
This parameter controls the non-commutativity of $\tE^a_i$ and $\phi^a_i$
(see \eqref{com-EE} and \eqref{com-phiphi}) and will play a crucial role in the following analysis.

It remains to analyze the stability conditions for the constraints $\pi$, $\CH$ and $\phi_{ij}:=\Et_{a(i}\phi^a_{j)}$.
They lead to equations involving Lagrange multipliers that can be written in the following form
\be
\CE\lapse= A_{ij}\lambda^{ij},
\qquad
\CE\rho= B_{ij}\lambda^{ij},
\label{relrhoNnew}
\ee
\be
\rho A_{ij}-\lapse B_{ij}={\eps_{(i}}^{kl}\lambda_{j)k}v_l,
\label{resLMnew}
\ee
where the phase space functions $\CE$, $A_{ij}$ and $B_{ij}$ are defined in \eqref{defCE} and \eqref{defAB},
and we introduced the vector
\be
v_i:=2\ab\tE^a_i\p_a\Lambda^{-1}.
\label{defv}
\ee
Note that taking the trace in the equation \eqref{resLMnew}, one arrives at a simple relation,
\be
\rho A=\lapse B,
\label{relAB}
\ee
where $A={A^i}_i$, $B={B^i}_i$. It requires that if $A$ vanishes then $B$ vanishes as well,
otherwise the lapse function would vanish and this contradicts the invertibility of the metric.
For the generic analysis we will assume that $A$ does not vanish. The case of both $A$ and $B$
vanishing is nevertheless physically interesting since it is realized for the de Sitter background
\eqref{deSit}. This special case will be analyzed in section~\ref{subsec-deSitter} below.

For non-vanishing $\CE$, the two equations \eqref{relrhoNnew} can be used to express $\lapse$ and $\rho$ in terms of $\lambda^{ij}$
and one remains with 6 equations \eqref{resLMnew} on these Lagrange multipliers.
However, despite the number of multipliers is equal to the number of equations, not all $\lambda^{ij}$ get fixed.
Indeed, contracting \eqref{resLMnew} with $\lambda^{ij}$, one obtains
\be
\rho A_{ij}\lambda^{ij}=\lapse B_{ij}\lambda^{ij},
\label{redunt}
\ee
which also follows directly from \eqref{relrhoNnew}. Thus, at least one equation out of 6 is redundant.
Furthermore, as we will see below, some of these stability conditions do not fix Lagrange multipliers,
but turn out to be secondary constraints.

It is instructive to compare the present situation to the
canonical analysis of GR in the first order formalism
(see e.g. \cite{Peldan:1993hi,Alexandrov:1998cu,Alexandrov:2000jw}).
In that case $\alpha=0$ and the constraint $\pi$ is missing
so that the above equations reduce to $\lambda^{ij}B_{ij}|_{\alpha=0}=\lapse B_{ij}|_{\alpha=0}=0$.
Since lapse is non-vanishing, this result implies the existence of 6 secondary constraints $B_{ij}|_{\alpha=0}\approx 0$,
which are in fact part of the vanishing torsion condition.

In our case the situation is more complicated. How many equations fix Lagrange multipliers and how many of them
generate secondary constraints depends on the properties of the $6\times 6$ matrix
\be
M_{ij}^{kl}:=\delta_{(i}^{(k}{\eps_{j)}}^{l)m}v_m
\label{matM}
\ee
appearing on the r.h.s. of \eqref{resLMnew} and multiplying $\lambda_{kl}$.
This matrix in turn is determined by the vector $v_i$ in \eqref{defv}.
There are two distinct cases, when this vector is zero or not, leading to different canonical structures.

The case of vanishing $v_i=0$ occurs either if $\ab=0$, or the field $\Lambda$ is spatially constant, $\p_a\Lambda=0$.
The first situation is a restriction on the parameters of the theory, the one anticipated in \eqref{condparam}.
The second situation is on the other hand a condition on a phase space variable, that
 identifies a class of backgrounds where the theory changes its canonical structure.
The presence of such special sectors in the phase space shows that our model is an example of irregular system.
Although the backgrounds with a spatially constant $\Lambda$
are interesting especially for cosmological applications \cite{Alexander:2020umk},
we will not analyze them in detail in this paper,
concentrating instead on {\it generic} backgrounds. We will thus only consider the case
$v_i=0$ obtained from $\ab=0$.

Since the analysis and the resulting canonical structure are different
in the two cases specified above, we consider them in the two consecutive subsections.

\subsection{Special case: scalar-tensor theories}
\label{case-spec}

We consider here the special case when $\ab=0$. This condition can be used to eliminate
$\b$ or equivalently $\a$ from the action, obtaining
\begin{align}
S(e,\om,\L) &=S_1(e,\om,\L) + \f\a2 \int \( \frac{1}{2}\,\eps_{IJKL} + \f{2\g}{\g^2-1}\d^{IJ}_{KL} \)
\frac{1}{\L}\, F^{IJ}(\omega)\w F^{KL}(\omega).
\end{align}
As we will see below, this class of theories propagates at most 3 degrees of freedom. It
includes the Euler theory for $1/\g=0$, and the theory with no Euler term and
arbitrary coupling for the Pontryagin term for $\a= 0$, $\g= 1$.

To complete the canonical analysis in this case, we set $v_i=0$ as implied by $\ab=0$, and observe that
the stability condition \eqref{resLMnew} reduces to
\be
\rho A_{ij}=\lapse B_{ij}.
\label{stabAB}
\ee
Expressing $\rho$ in terms of the lapse using the relation \eqref{relAB} and taking into account that the lapse is non-vanishing,
one arrives at  equations which do not involve any Lagrange multipliers and must therefore be interpreted as secondary constraints,
\be
\psi_{ij}= B A_{ij}-A B_{ij}\approx 0.
\label{defpsi-1}
\ee
Since the trace ${\psi^i}_i$ vanishes identically, there are only 5 independent secondary constraints, which is related
to the redundance noticed around \eqref{redunt}.

Thus, our next problem is to study the stability of the secondary constraints \eqref{defpsi-1}.
To this end, let us take into account the conditions on the Lagrange multipliers found above.
These conditions include: vanishing of the antisymmetric part of $\lambda^{ij}$, the expression \eqref{fixm} for $m_i$,
and the expressions of $\rho$ and $\lapse$ through $\lambda^{ij}$ resulting from \eqref{relrhoNnew}.
Alternatively, we can split $\lambda^{ij}$ into two parts: $\lambda_A=A_{ij}\lambda^{ij}$ and
\be
\hlam^{ij}=\lambda^{ij}-\delta^{ij} A^{-1}A_{kl}\lambda^{kl}
\ee
satisfying $A_{ij}\hlam^{ij}=0$. Then, instead of expressing $\lapse$ through
$\lambda^{ij}$, one can express $\lambda_A$ through $\lapse$
so that the set of independent Lagrange multipliers will now consist of $\hlam^{ij}$ and the lapse $\lapse$.
The advantage of this way of proceeding is that we do not have to require that $\CE$ is non-vanishing and
we can keep the geometric field $\lapse$ as an independent variable.
Substituting all conditions into the total Hamiltonian \eqref{Htotnew}, one finds
\be
-\Htot=\CR(n)+\CD(\vec\shift)+\hCH(\lapse)+\hphi(\hlam),
\label{Htot2}
\ee
where
\bea
\hCH&=&\CH+BA^{-1}\pi +\CE A^{-1} {\phi^k}_k+\(BA^{-1}\CA_i+\CB^-_i\)\CK^i,
\label{defPhi0}
\\
\hphi_{ij}&=& \phi_{ij}-A_{ij}A^{-1} {\phi^k}_k.
\label{defhPhi}
\eea

With the help of the commutation relations provided in Appendix~\ref{ap-useful} (see \eqref{commA} and \eqref{commB}),
it is easy to check that $\psi_{ij}$ are weakly commuting with the diffeomorphism and rotation constraints
\bea
\{ \CD(\vec\shift), \psi_{ij}\} &=&\p_a (\shift^a\psi_{ij}),
\label{com-Dpsi}
\\
\{ \CR(n), \psi_{ij}\} &=& 2{\eps_{(ik}}^l n^k \psi_{lj)}.
\label{com-Rpsi}
\eea
Therefore, the stabilization of $\psi_{ij}$ leads to 5 equations
\be
\{\psi_{ij},\hCH(\lapse)\} + \{\psi_{ij},\hphi(\hlam)\}=0.
\label{stabpsi}
\ee
The second term may involve in general derivatives on $\hlam^{ij}$, coming from
the commutation relations between $\phi_{ij}$ and $A_{ij}$, $B_{ij}$, from which the secondary constraints
are constructed. However, a look at these commutators, given in \eqref{com-phiA} and \eqref{com-phiB},
shows that there is only one term with a derivative acting on $\lambda_{ij}$,
and this term is proportional to $\ab$ and thus absent in the special case being treated here.
As a consequence, we can write the second term as
\be
\{\psi_{ij},\hphi(\hlam)\}=\Delta_{ij,kl}\hlam^{kl},
\label{defDel}
\ee
where $\Delta_{ij,kl}$ is a matrix valued function and not
a differential operator.
Furthermore, on a generic background the $5\times 5$ matrix $\Delta_{ij,kl}$ is invertible\footnote{It involves
the contribution which coincides with the commutator of the primary simplicity and secondary constraints for GR
whose invertibility is ensured by the invertibility of the tetrad. }
and hence the equation \eqref{stabpsi}
can be solved with respect to $\hlam^{ij}$, giving
\be
\hlam^{ij}=-(\Delta^{-1})^{ij,kl}\{\psi_{kl},\hCH(\lapse)\}.
\label{reshlam}
\ee
Thus, the stability condition of the secondary constraints does not produce any further constraints and the stabilization procedure
stops at this point.

Since the lapse $\lapse$ remained unfixed, the combination of constraints which it multiplies in the total Hamiltonian is
ensured to be first class and, in fact, it is this combination that plays the role of the Hamiltonian constraint in this formalism.
From \eqref{reshlam}, it is easy to see that it is given by
\be
\crH:= \hCH-\{\psi(\hphi\,\Delta^{-1}),\hCH\}.
\label{finalH}
\ee
This constraint commutes with itself only weakly, unlike the original constraint $\CH$ (see \eqref{com-HH}).
In particular, the commutator contains the standard contribution given by the vector constraint
$-\CV(\vec K_{N,M})$ where $K_{N,M}^a= q q^{ab}(\Nt\p_b\Mt-\Mt\p_b\Nt)$,
which is characteristic of the diffeomorphism algebra (cf. \eqref{defHH-SD}).
The latter appears as follows: the first term in $\CB^-_i$
\eqref{defCBnew} gives rise to the contribution $-\tE^a_i\p_a\CK^i$ in \eqref{defPhi0},
which then produces the vector constraint through the commutation relation \eqref{com-cKH}.

Summarizing the results of our analysis, we have 7 first class constraints $\CR_i$, $\CD_a$ and $\crH$
corresponding to the unfixed gauged symmetries of the model,
together with 18 second class constraint $\CK_i$, $\phi^a_i$, $\pi$ and $\psi_{ij}$.
These constraints are imposed on a $38$ dimensional phase space.
This means that the reduced phase space has dimension $38-2\cdot 7-18=6$ and corresponds
to 3 propagating degrees of freedom. As we already discussed, the new degrees of freedom
of these theories reside in torsion.
The counting suggests that we are looking
at a single scalar mode of torsion, which could be described by taking either
of its spin-1 parts (the trace or the completely antisymmetric irreps)
to be given by the gradient of a scalar. It might be possible to view
this special case as a type of scalar-tensor theory,
which is why we refer to the special case in these terms in the title of this section.

The counting we obtained can be compared with older work appeared in  \cite{Mignemi:1998us}.
This paper studied the canonical analysis of a MacDowell-Mansouri-type of action \cite{MacDowell:1977jt} with
de Sitter (or anti-de Sitter)
gauge symmetry
group and involving an additional scalar field in the fundamental representation. This model turns out to be equivalent
to the Euler theory, which is included in the special case treated here.\footnote{The equivalence is established
in a similar way as the original MacDowell-Mansouri theory is shown to be equivalent to general relativity:
one introduces a partial gauge fixing of the larger internal group,
which requires the scalar multiplet to have only one non-vanishing component and thereby breaks
the symmetry down to the Lorentz subgroup,
and performs a field rescaling.}
The result claimed in \cite{Mignemi:1998us} is that such model has 6 degrees of freedom,
which is in contradiction to what we find.
However, the 6 degrees of freedom are counted assuming that a certain matrix does not have additional
null eigenvectors beyond a few ones explicitly found in the paper.
We believe that this assumption is unjustified, and that additional null eigenvectors exist and bring the counting to match ours.
It would be interesting to fill this gap and establish a direct link between the canonical structures of the two models.

\subsection{Generic case}
\label{case-gen}

Let us now consider the case of non-vanishing $v_i$.
To disentangle the corresponding stability conditions, note
that the $6\times 6$ matrix $M_{ij}^{kl}$ \eqref{matM} has two eigenvectors with vanishing eigenvalues,
$\delta_{kl}$ and $v_kv_l$, and hence its rank is equal to 4.
This implies that 6 stability conditions \eqref{resLMnew} split into 3 groups of equations.
First, taking the trace, as was already noticed, leads to the relation \eqref{relAB}.
Second, contracting \eqref{resLMnew} with $v^iv^j$ and using \eqref{relAB},
one generates one secondary constraint of the following form
\be
\psi_v= \frac{v^iv^i}{v^2}\,\psi_{ij}=BA_v-AB_v\approx 0,
\label{secondpsi}
\ee
where $A_v=v^iv^j A_{ij}/v^2$, $B_v=v^iv^j B_{ij}/v^2$.
Finally, the remaining 4 equations can be used to express 4 components of $\lambda^{ij}$, orthogonal to $\lambda={\lambda^k}_k$
and $\lambda_v=v_iv_j\lambda^{ij}/v^2$, through $\rho$ and $\lapse$.

To do the last step explicitly, let us introduce the projector on the space of traceless symmetric matrices orthogonal to $v_kv_l$.
It is easy to check that it is given by
\be
\CI_{ij}^{kl}(v)= \delta_{(i}^k\delta_{j)}^l-\hf\, \delta_{ij}\delta^{kl}+\frac{1}{2v^2} \, \delta_{ij}v^kv^l
+\frac{1}{2v^2} \, \delta^{kl}v_i v_j-\frac{3}{2(v^2)^2} \, v_i v_j v^kv^l.
\ee
Then the matrix
\be
\begin{split}
N_{ij}^{kl}=&\, -\frac{1}{v^2}\, M_{ij}^{kl}-\frac{6v_{(i} v^m}{(v^2)^2}\, M_{mj)}^{kl}
= -\(\frac{1}{v^2}\,\delta_{(i}^{(k}+\frac{3v_{(i}v^{(k}}{(v^2)^2}\){\eps_{j)}}^{l)m}v_m
\end{split}
\label{resinvmat}
\ee
satisfies
\be
N_{ij}^{mn}M_{mn}^{kl}=\CI_{ij}^{kl}(v),
\ee
which means that it can be considered as the inverse to $M_{ij}^{kl}$ on the subspace we are interested in.
Using this result and substituting the decomposition
\be
\lambda^{ij}=\hlam^{ij}+\(\delta_{(i}^k\delta_{j)}^l-\CI_{kl}^{ij}(v)\)\lambda^{kl}
=\hlam^{ij}+\hf\, \delta^{ij}\lamp-\frac{v^iv^j}{2v^2}\,\lamm,
\label{decomp-lam}
\ee
where
\be
\hlam^{ij}:=\CI_{ij}^{kl}(v)\lambda_{kl},
\qquad
\lamp:=\lambda-\lambda_v,
\qquad
\lamm:=\lambda-3\lambda_v,
\ee
into \eqref{resLMnew},
one finds that $\lampm$ drop out, whereas $\hlam_{ij}$ can be found to be
\be
\hlam_{ij}
= N_{ij}^{kl}\(\rho A_{kl}-\lapse B_{kl}\)
=\lapse A^{-1} N_{ij}^{kl}\(B A_{kl}-A B_{kl}\),
\label{solhlam}
\ee
where in the last equality we used \eqref{relAB}.

It remains to analyze the two stability conditions \eqref{relrhoNnew}.
In the parametrization \eqref{decomp-lam} they can be rewritten as a system of linear equations
\be
\begin{split}
A\lamp-A_v\lamm=&\, 2\CE\lapse- 2A_{ij}\hlam^{ij},
\\
A(B\lamp-B_v\lamm)=&\, 2\CE B\lapse- 2A B_{ij}\hlam^{ij}.
\end{split}
\label{relrhoNlam}
\ee
Using the constraint \eqref{secondpsi}, the second equation is easily shown to be equivalent to the first one, provided
\be
(B A_{ij}-A B_{ij})\hlam^{ij}=0.
\ee
This is indeed the case, as it follows either from \eqref{resLMnew}
by contracting it with $\hlam^{ij}$, or directly form \eqref{solhlam} due to
the antisymmetry of the matrix \eqref{resinvmat}. This is the same redundancy that has been noticed in \eqref{redunt}.
Hence, there is only one equation relating three Lagrange multipliers,
\be
A\lamp-A_v\lamm=
2\(\CE+A_{ij} N^{ij,kl}B_{kl}\)\lapse.
\label{constr-lam}
\ee

Solving \eqref{constr-lam} with respect to $\lamp$, one concludes that
the first step of the stabilization procedure gives one secondary constraint $\psi_v$ and leaves
unfixed the following Lagrange multipliers: $n^i$, $\shift^a$, $\lapse$
and $\lamm$. Taking into account all conditions on the Lagrange multipliers,
the total Hamiltonian \eqref{Htotnew} takes the form
\be
-\Htot=\CR(n)+\CD(\vec\shift)+\hCH(\lapse)+\hphi_v(\lamm),
\label{Htot2-gen-al}
\ee
where
\bea
\hCH&=& \CH+A^{-1}B\pi
+A^{-1} N^{ij,kl}\(B A_{kl}-A B_{kl}\)\(\phi_{ij}+\frac{v_i\CK_j}{2q\(1+1/\gamma^{2}\)}\)
\nn\\
&&
+A^{-1}\(\CE+A_{ij} N^{ij,kl}B_{kl}\)\(\phi-\frac{v^m \CK_m}{q\(1+1/\gamma^{2}\)}\)
+\(A^{-1}B\CA_i+\CB^-_i\)\CK^i,
\\
\hphi_v&=& \frac{A_v\phi-A \phi_v}{2A}
-\frac{A_v v^i \CK_i}{2qA\(1+1/\gamma^{2}\)}\,,
\eea
we defined $\phi:={\phi^k}_k$, $\phi_v:=v^iv^j\phi_{ij}/v^2$ and also used that (see \eqref{fixm})
\be
m_i= \(A^{-1}B\CA_i +\CB^+_i\)\lapse+\frac{\hlam_{ij}v^j-v_i \lamp}{2q\(1+1/\gamma^{2}\)}.
\label{m-new}
\ee

We still have to stabilize the secondary constraint \eqref{secondpsi}.
The analysis is very similar to the previous case.
First, from \eqref{com-Dpsi}, \eqref{com-Rpsi} and \eqref{com-Dv}, one finds that $\psi_v$
is weakly commuting with the diffeomorphism and rotation constraints
\bea
\{ \CD(\shift), \psi_v\} &=&\p_a (\shift^a\psi_v),
\\
\{ \CR(n), \psi_v\} &=& 0.
\eea
As a result, the stability condition of $\psi_v$ reads
\be
\{\hCH(\lapse)+\hphi_v(\lamm),\psi_v\}=0.
\label{stabpsi-gen}
\ee
Importantly, the Lagrange multiplier $\lamm$ enters this equation without derivatives.
Indeed, since the commutators of the boost constraint $\CK(m)$ with $A_{ij}$ and $B_{ij}$, \eqref{com-KA} and \eqref{com-KB},
do not involve derivatives on $m_i$, the only commutator we have to analyze is
\be
\{A_v\phi(\lamm)-A \phi_v(\lamm),BA_v-AB_v\}.
\ee
The only terms with a derivative on $\lamm$ come from the commutator \eqref{com-phiA} and are encoded by the contribution
\be
\frac{2\ab}{q\Lambda^2}\, {\eps_{(i}}^{kl}\tE^a_k\p_a\lambda_{j)l}.
\label{dangcom}
\ee
However, it is easy to see that it vanishes provided $\lambda_{jl}$ is replaced by $\delta_{jl}\lamm$ or $v_lv_j\lamm$
simultaneously with contracting it with $\delta_{ij}$ or $v_iv_j$.
This ensures that $\lamm$ appears in the stability condition \eqref{stabpsi-gen} without derivatives.
As a result, this condition can be used to express this Lagrange multiplier as a linear function of the lapse,
so that the first class Hamiltonian constraint can be written as
\be
\crH(\lapse)= \hCH(\lapse)-\frac{\{\hCH(\lapse),\psi_v\}}{\{\hphi_v,\psi_v\}}\,\hphi_v.
\ee
This step finishes the stabilization procedure.

Thus, in total we have again 7 first class constraints $\CR_i$, $\CD_a$ and $\crH$, but in contrast to the previous case
they are supplemented by only 14 second class constraints $\CK_i$, $\phi^a_i$, $\pi$, $\psi_v$.
Therefore, the reduced phase space has dimension $38-2\cdot 7-14=10$ and corresponds
to 5 propagating degrees of freedom.

\subsection{De Sitter background}
\label{subsec-deSitter}

The previous construction assumed no particular relations satisfied by the fields
except the constraints and stability conditions found
in the process of canonical analysis.
Now let us instead consider the special case of the de Sitter background \eqref{deSit},
which is a solution when the parameters are restricted to satisfy \eqref{deSit-param}.

This background requires a special attention, because it violates some of the assumptions
made in the previous sections. This can be easily seen in the time gauge,
for which the de Sitter condition \eqref{deSit} gives
\be
F_{ab}^{0i}=0,
\qquad
F_{ab}^i=\frac{\Lambda}{3}\, \epst_{abc}\tE^c_i.
\label{deSit-cond}
\ee
Using these values in the functions entering the momenta \eqref{momenta}, one finds
\be
\tcP^a_i(\omega)=-\tE^a_i,
\qquad
\tcQ^a_i(\omega)=-\frac{1}{\gamma}\, \tE^a_i.
\ee
This means that background values of momenta $\tP^a_i$ and $\tQ^a_i$ identically vanish.
Furthermore, the formula for the torsion \eqref{deSit-tor} implies
\be
\nabla_{[a} E_{b]}^i=\hf\, E_{[a}^i\p_{b]}\log\Lambda,
\ee
where $\nabla_{a}$ is the $SU(2)$ covariant derivative defined in \eqref{defnabla}.
Then it is easy to check that the following quantities appearing in various stability conditions all vanish
\be
\CA_i=A_{ij}=B_{ij}=\cE=0.
\label{zero-deSit}
\ee
This makes the above analysis inapplicable to the de Sitter background because
it was based on the assumption of non-vanishing $A$.
Hence, it should be redone and the canonical structure may change.

Before proceeding, let us clarify what it means to perform the canonical analysis on a fixed background.
The only meaningful way to do this is to expand the fields around their background values.
If one keeps all orders of the expansion, one effectively considers the theory on a generic background,
just written in a shifted variables.
If however one keeps only linear terms in equations of motion (hence quadratic in the action),
this is not true anymore and one may find something new.

Thus, our goal will be to understand the canonical structure of the {\it linearized} theory obtained
by expanding the fields around a solution satisfying \eqref{deSit}.
The corresponding action has already been obtained in \eqref{Slinear}.
However, instead of performing the analysis from scratch, we can use our previous results.
Let us write the initial Hamiltonian (in our case this is \eqref{Htotnew}) as $-\Htot=\zeta^\alpha\Phi_\alpha$
where $\Phi_\alpha$ are primary constraints and $\zeta^\alpha$ are their Lagrange multipliers.
Then the Hamiltonian describing the fluctuations
(which we denote by an upper index indicating the perturbation order)
is obtained by expanding $\Htot$ to quadratic order which gives
\be
-\Htot^{(2)}= \zetaqi{\alpha}\Phiq_\alpha+\zeta^\alpha\Phi^{(2)}_\alpha.
\label{Hexp}
\ee
In contrast to the original Hamiltonian, $\Htot^{(2)}$
is not weakly vanishing due to background values of the Lagrange multipliers
(which at least for the lapse $\lapse$ is guaranteed to be non-zero).
More importantly is that the constraints, which appear as functions $\Phiq_\alpha$ multiplied by
fluctuation of the Lagrange multipliers, are linear in fluctuations of the canonical fields.
Therefore, their stability conditions can be obtained by simply linearizing the non-perturbative stability conditions.

To do the linearization explicitly, one should take into account
that the background value of the Lagrange multiplier $\rho$ is zero,
whereas $\lambda_a^i$ can be found from Hamilton's equations of motion
for the combination of connection components that is conjugate
to the combination of momenta entering the constraint \eqref{primcon}.
More precisely, one has
\be
\begin{split}
\lambda_a^i=&\, -\p_t\(\hf\, {\eps^i}_{jk}\omega_a^{jk}-\frac{1}{\gamma}\, \omega_a^{0i}\)+
D^{(\omega)}_a\(\hf\, {\eps^i}_{jk}\omega_0^{jk}-\frac{1}{\gamma}\, \omega_0^{0i}\)
\\
=&\, -\hf\, {\eps^i}_{jk}F_{0a}^{jk}+\frac{1}{\gamma}\,F_{0a}^{0i}.
\end{split}
\ee
This can be further  expressed through the tetrad and $\Lambda$ using the de Sitter condition \eqref{deSit}.
Below we will need the expression only for the symmetric part $\l^{ij}$ of the Lagrange multipliers which turns out to be given by
\be
\lambda^{ij}= \frac{\lapse\Lambda \,\delta^{ij}}{3\gamma (1+1/\gamma^2)}.
\label{backgr-lam}
\ee

Linearizing the stability conditions for primary constraints, one then finds that
the Lagrange multipliers $\mq_i$ are fixed by the linearized version of \eqref{fixm},
whereas thanks to the vanishing results \eqref{zero-deSit},
the stability conditions \eqref{relrhoNnew} and \eqref{resLMnew} reduce to
\be
\lapse \CEq =\lambda^{ij}\Aq_{ij},
\qquad
\lambda^{ij}\Bq_{ij}=0,
\qquad
\lapse \Bq_{ij}=M_{ij}^{kl}\lamq_{kl},
\label{linearstab}
\ee
where $M_{ij}^{kl}$ is the matrix \eqref{matM}.
Taking into account \eqref{backgr-lam}, the first two equations
seem to give rise to two secondary constraints
\be
\Psiq=\CEq-\frac{\Lambda\,\Aqi{i}_i}{3\gamma (1+1/\gamma^2)}\approx 0,
\qquad
\psiq={\Bqi{i}_i}\approx 0.
\label{newIIclconstr}
\ee
However, the first is not a new constraint, because it is straightforward to verify that
\be
\Psiq= \Lambda^{-1} \CHq.
\label{relEH}
\ee
Hence, we have only one secondary constraint $\psiq$.
Finally, the nature of the last equation in \eqref{linearstab} depends on the vector $v_i$.
Thus, as before, we have to consider two cases, when $v_i$ is zero or not, separately.

In the special case $v_i=0$, which corresponds to $\gamma=\infty$ or spatially constant $\Lambda$,
the last stability condition generates 6 secondary constraints
\be
\psiq_{ij}=\Bq_{ij}\approx 0,
\ee
of which $\psiq$ \eqref{newIIclconstr} is a linear combination.
These constraints can be viewed as a generalization of the linearized secondary constraints of GR
in the first order formalism. As a consequence, they do not commute
with 6 primary constraints $\phiq_{ij}$, so that these 12 constraints form second class pairs.
On the other hand, evaluated on the de Sitter background, all commutation relations of $\piq$ with other constraints
can be checked to be zero. Therefore, in contrast to the canonical analysis on generic backgrounds, it is of first class.
In summary, we have found 8 first class constraints $\CRq_i$, $\CDq_a$, $\crHq$, $\piq$,
and 18 second class constraints $\CKq_i$, $\phiqi{a}_i$, $\psiq_{ij}$.
Thus, the dimension of the reduced phase space is equal to $38-2\cdot 8-18=4$, which corresponds
to 2 propagating degrees of freedom.

For non-vanishing $v_i$, the analysis proceeds in a similar way.
Since in this case the matrix $M_{ij}^{kl}$ has rank 4, the last equation in \eqref{linearstab}
fixes 4 Lagrange multipliers $\lamq_{ij}$
and gives rise to two secondary constraints obtained by taking the trace and contracting with $v_i$,
One of them coincides with $\psiq$ in \eqref{newIIclconstr}, and the second is
\be
\psiq_v=\frac{v^i v^j}{v^2}\, \Bq_{ij}\approx 0.
\ee
The two constraints $\psiq$ and $\psiq_v$ do not commute with the two remaining components of the primary constraints
$\phiq_{ij}$, i.e. $\phiq$ and $\phiq_v$,
whereas $\piq$ remains commuting with all constraints as in the previous case.
Thus, we end up having 8 first class constraints $\CRq_i$, $\CDq_a$, $\crHq$, $\piq$,
and 14 second class constraints $\CKq_i$, $\phiqi{a}_i$, $\psiq$, $\psiq_v$.
Therefore, the reduced phase space has dimension $38-2\cdot 8-14=8$ and corresponds
to 4 propagating degrees of freedom.

We observe that in both cases the difference in the resulting canonical structure comparing to the one on a generic background
is that one acquires an additional second class constraint $\psiq$, while the constraint $\piq$,
which was second class, is converted now to first class.
As a result, one looses one degree of freedom.

The conversion of $\piq$ into first class means that it must be the generator of a gauge symmetry.
This symmetry is nothing but the conformal symmetry discussed in section \ref{deSitter-sol}.
Indeed, it is easy to calculate the following commutation relations on the de Sitter background
\be
\{\piq(\rho), \omqi{IJ}_a\}=0,
\qquad
\{\piq(\rho), \tEqi{a}_i\}=\frac{\rho}{\Lambda}\, \tE^a_i,
\qquad
\{\piq(\rho), \Lamq\}=-\rho .
\ee
The second commutator implies
\be
\{\piq(\rho), \eqi{i}_a\}=\frac{\rho}{2\Lambda}\, e_a^i.
\ee
After identification $\rho=2\Lambda\veps$, these commutation relations reproduce the infinitesimal
conformal transformation \eqref{conftr-q}.

\section{Self-dual theory}
\label{SecSD}

It is a well-known fact that GR in tetrad variables admits a self-dual formulation
which can be described using the Einstein-Cartan-Holst action with  $\g=\pm\I$.
At these values of the Immirzi parameter, the action depends only on (anti-)self-dual part of
the connection which makes the theory very special.
For instance, being holomorphic it realizes the full Lorentz symmetry group as $SO(3,1)\simeq SO(3,\C)$ and
leads to a canonical formulation identical to the one of the real theory
in the time gauge without actually imposing any gauge restriction.
It is this complex Ashtekar formulation \cite{Ashtekar:1986yd,Ashtekar:1987gu}
that was the starting point of the loop approach to quantization of GR \cite{Rovelli:2004tv,Rovelli:1987df,Rovelli:1989za},
while its related covariant Plebanski formulation \cite{Plebanski:1977zz} underlies the spin foam approach
\cite{Perez:2003vx,Alexandrov:2011ab}.
More recently, a whole class of modified theories of gravity describing the dynamics of the self-dual connection
has been proposed, which is distinguished by the property that, similarly to GR, these models propagate only two
(holomorphic) degrees of freedom \cite{Krasnov:2007cq,Krasnov:2008fm,Krasnov:2009iy}. Furthermore, a self-dual
formulation was shown to exist even for ghost free massive gravity \cite{Alexandrov:2012yv}
where it also considerably simplifies the analysis.

In the modified theory of gravity considered here, fixing $\gamma=\pm \I$ is not enough yet to get a self-dual
theory due to the presence of the second term $S_2$ \eqref{defL2}.
However, this can be achieved by fixing also $\gamma'$ to the same value.
Therefore, given the special role played by holomorphic formulations of GR and its modifications,
in this section we study our model specialized for
\be\label{SDpar}
\g=\g'=-\I,
\ee
which selects its self-dual sector.
It is particularly interesting because it is this self-dual model that
appeared in the original proposal of \cite{Alexander:2018djy}
and suggested a new interpretation of the standard Kodama state \cite{Kodama:1990sc} as a transition amplitude.
Therefore, a proper canonical analysis of this case appears to be important
in view of possible applications of this model to quantum gravity.

The canonical analysis of the self-dual theory cannot simply be deduced from the general analysis presented
in the previous section. This is because the
expression \eqref{defE2} for the triad becomes singular in the limit \eqref{SDpar}.
Even if one uses the freedom to change this definition discussed above \eqref{defE2}
(that is, by adding terms proportional to the primary constraint \eqref{primcon})
at the price of having more complicated commutation relations and hence the stability analysis,
the analysis presented above fails for such $\gamma$ because several crucial
commutators become vanishing (e.g. \eqref{com-Kphi}).
Therefore we start from scratch.

Substituting \eqref{SDpar} into the two actions \eqref{twoactions} and noting that $\opP(-\I) = 2\I \opP_+$ where
$
\opP_{\pm} =\f12(\Id \mp \I\star)
$
are the self-dual and anti-self-dual projectors, respectively, one finds that
\begin{subequations}
\bea
S_1 &=&\int \( \Si_i\w F^i(A)+\f{\I\L}{6}\,\Si_i\w \Si^i\),
\label{defL1-sd}
\\
S_2 &=&\I\a \int\L^{-1} F_i(A)\w F^i(A),
\label{defL2-sd}
\eea
\label{twoactions-sd}
\end{subequations}
where we defined the self-dual connection and Plebanski 2-form
\be
\begin{split}
A^i=&\, -2\I \opP_+{}^{0i}_{KL}\om^{KL} =\hf\,\eps^i_{jk}\om^{jk} -\I\om^{0i},
\\
\Si^i =&\,  2 \opP_+{}^{0i}_{IJ} e^I\w e^J = e^0\w e^i + \f{\I}{2}\, \eps^{ijk} e^j\w e^k,
\end{split}
\label{defSig}
\ee
whereas $F^i(A)=\de A^i-\f12\,\eps^{ijk} A^j\w A^k$ is the self-dual curvature.

An interesting observation is that, upon integration by parts, the action \eqref{defL2-sd}
can be rewritten in terms of the Chern-Simons Lagrangian
\be
\CL_{\scr CS}(A) = A_idA^i-\f13\,\eps_{ijk}A^i\w A^j\w A^k.
\label{CS}
\ee
This follows from the identity $ \de \CL_{\scr CS}=F_i\w F^i$ which implies that
\be
S_2 =-\I\a \int\de\L^{-1}\w \CL_{\scr CS}(A).
\label{defL2-sdCS}
\ee
It is this observation that was the basis for the above mentioned interpretation of the Kodama state,
which is constructed as the exponential of the CS Lagrangian.

Our goal is to perform the canonical analysis of this model.
To this end, both forms of $S_2$, \eqref{defL2-sd} and \eqref{defL2-sdCS}, can be used as the starting point.
The difference between the resulting canonical structures should be captured merely by a canonical transformation.
To present a new look at the model and given the important role played by the CS Lagrangian,
we choose to start with the action \eqref{defL2-sdCS}.

\subsection{Canonical analysis}
\label{subsec-canan-sf}

The first step is, as usual, to introduce a convenient parametrization of the fields adapted to the canonical analysis.
Such parametrization can be obtained by plugging the ansatz \eqref{dec-tetrad}
into the definition of the Plebanski 2-form \eqref{defSig}
and suggests to define
\be
\tSig^a_i= \hf\,\teps^{abc}\Sigma_{bc}^i,
\qquad
\Sigma_{0a}^i= -\frac{1}{2}\,\lapse\epst_{abc}\eps^{ijk} \tSig^b_j\tSig^c_k-\epst_{abc}\shift^b\tSig^c_i.
\label{deftSig}
\ee
It is easy to check that the determinant of the induced metric \eqref{indmet}
written in terms of these fields is given by
\be
q=\frac{\I}{6}\,\epst_{abc}\eps^{ijk}\tSig^a_i\tSig^b_j\tSig^c_k.
\ee
Since both $A^i$ and $\Sigma^i$ are complex fields and the action is holomorphic,
it natural to regard $\Lambda$ as complex field as well.
Then we find it convenient to redefine this scalar field as
\be
\vp=-i\L^{-1}.
\label{defvp}
\ee

In terms of these fields, the action of the self-dual model composed of \eqref{defL1-sd} and \eqref{defL2-sdCS}
takes the following form
\be
S=\int \de t \de^3\xc\left[\tP^a_i \dot A_a^i
+ \a \dot\vp\, \CL_{\scr CS}+A_0^i \CG_i + \shift^a\CV_a + \lapse\CH \right],
\label{SDact-decomp}
\ee
where
\be
\tP^a_i = \tSig^a_i -\a\tl\eps^{abc}A_{bi}\p_c\vp
\ee
is the momentum conjugate to the self-dual connection, whereas
\begin{subequations}
\bea
\CG_i &=& D_a\tSig^a_i -\a\tl\eps^{abc}F_{ab}^i\p_c\vp,
\label{Gauss-sf}
\\
\CV_a &=& -\tSig^b_i F^i_{ab},
\label{CV-sf}
\\
\CH &=& -\frac{1}{2}\,\eps^{ijk}\tSig^a_i \tSig^b_j F^k_{ab}+\frac{\I q}{\vp}\, ,
\label{CH-sf}
\eea
\label{primeconstr-SD}
\end{subequations}
with $D_a$ being the covariant derivative $D_av_i=\p_av_i-{\eps_{ij}}^kA_a^j v^k$, are primary constraints
generating gauge transformations, spatial and time diffeomorphisms, respectively.
Denoting $\pi$ the momentum conjugate to $\vp$, we have the following canonical Poisson brackets
\be
\{A_a^i(\xc),\tP^b_j(\yc)\} = \d_a^b\d^i_j\,\d^{(3)}(x,y),
\qquad
\{ \vp(\xc),\pi(\yc)\}=\delta^{(3)}(\xc,\yc).
\label{cancom-sf}
\ee
The momentum $\pi$ however is not an independent quantity. As follows from \eqref{SDact-decomp},
it satisfies an additional primary constraint
\be
\Pi=\pi-\a\CL_{\scr CS}(A)\approx 0.
\ee
Thus, instead of 20 primary constraints of the general case, the self-dual theory has only 8 primary constraints,
though considered as holomorphic equations.
Finally, we rewrite the Gauss constraint in terms of the canonical momenta and redefine the vector constraint $\CV_a$
similarly to \eqref{constrD}
\begin{subequations}
\bea
\CG_i&=& D_a\tP^a_i -\a\teps^{abc}\p_a\vp\p_b A_c^i.
\label{Gauss-SD}
\\
\CD_a& =&  \CV_a+A_a^i\CG_i-\Pi\p_a\vp
\nn \\
&=& -\tP^b_i\p_a A_b^i+\p_b(A_a^i\tP^b_i)-\pi\p_a\vp.
\label{diff-SD}
\eea
\label{primeGD-SD}
\end{subequations}
\indent
The stability conditions of the primary constraints are obtained by commuting them with the total Hamiltonian
\be
-\Htot=\CG(n)+\CD(\vec\shift)+\CH(\lapse)+\Pi(\rho).
\label{Htot-SD}
\ee
Since the constraint algebra presented in appendix \ref{ap-conalg2} has only one non-vanishing commutator, namely,
the one of $\Pi$ with the Hamiltonian constraint $\CH$ \eqref{com-PiH}, there are only two non-trivial
stability conditions. Since the lapse is non-vanishing, they both give rise to one secondary
constraint
\be
\Psi=(1+2\alpha)\,\frac{q}{\vp^2}-\I\alpha\teps^{abc}\eps_{ijk} F_{ab}^i \tSig^d_j F_{cd}^k\approx 0.
\label{defPsi}
\ee

Thus, to complete the stabilization procedure, we have to study the stability condition of the new constraint.
To this end, we compute the following commutation relations
\begin{subequations}
\bea
\{\CG(n),\Psi\}&=& 0,
\\
\{ \CD(\vec N), \Psi\} &=& N^a\p_a \Psi+\Psi \p_a N^a,
\\
\{ \Pi(\rho),\Psi\} &=& \rho \[\frac{2(1+2\alpha)}{\vp^3}\,\bigl(\I\alpha\vp\cH+(1+\alpha)q\bigr)
+\I\alpha^2\eps_{ijk}\teps^{abc}F^i_{bc}\teps^{dgf}F_{gf}^j F_{ad}^k\],
\label{com-PiPsi}
\eea
\end{subequations}
where we used
\be
\{\tSig(f),\tSig(g)\}=2\alpha\int \de^3 \xc\,\teps^{abc}f_a^i g_b^i\p_c\vp.
\ee
Importantly, the commutator \eqref{com-PiPsi} does not contain derivatives acting on $\rho$. As a result, the stability
condition of $\Psi$, which reads as
\be
\{\CH(\lapse)+\Pi(\rho),\Psi\}=0,
\label{stabPsi}
\ee
can be solved with respect to this Lagrange multiplier expressing it in terms of the lapse.
As a result, we arrive to the conclusion that $\CG_i$, $\CD_a$ and
\be
\crH(\lapse)= \CH(\lapse)-\frac{\{\CH(\lapse),\Psi\}}{\{\Pi,\Psi\}}\,\Pi
\ee
are first class constraints, whereas $\Pi$ and $\Psi$ are second class.
Imposing these constraints on the 20-dimensional (complex) phase space, one gets the reduced phase space of dimension
$20-2\cdot 7-2=4$, i.e. 2 (holomorphic) degrees of freedom.
Thus, the canonical structure of the self-dual theory drastically differs from the one of the real model.

\subsection{De Sitter background}\label{sec-dSSF}

As we learnt in section \ref{subsec-deSitter},
the theory around the de Sitter background has a canonical structure different from that of a generic point in phase space.
This is true also in the self-dual theory.
Indeed, it is easy to see that the r.h.s. of the commutator \eqref{com-PiPsi}
vanishes on the de Sitter background so that the previous conclusion that $\Pi$ and $\Psi$ are second class fails.
Furthermore, it is natural to expect that one of the constraints should become first class as in the real theory,
and generate the additional gauge symmetry of the linearized perturbations around these backgrounds.
This expectation is confirmed proceeding as explained in section \ref{subsec-deSitter}. One finds that
(the linearized constraint) $\Pi^{(1)}$ becomes first class, and that
\be\label{protasov}
\Psiq=\frac{2\I}{\vp}\, \CHq
\ee
coincides with the linearized Hamiltonian constraint \eqref{CH-sf}.
This is a direct analogue of the relation \eqref{relEH} in the real case which similarly ensures the absence
of an additional secondary constraint.

As a result, on the de Sitter background we remain with 8 first class constraints $\CG_i$, $\CD_a$, $\CH$ and $\Pi$,
which lead to $20-2\cdot 8-4$ dimensional reduced phase space, i.e. 2 degrees of freedom.
Thus, the difference with the canonical structure on generic backgrounds
is that instead of two second class constraints one has now
only one, $\Pi$, which is first class. As in section \ref{subsec-deSitter},
one can check that it is the generator of the gauge transformation \eqref{conftr-q}.
Note that in contrast with the real theory, the extra gauge symmetry around de Sitter solutions
does not lead to a reduction of degrees of freedom. The reason for this is that
while $\Pi$ is converted into first class, $\Psi$ drops out.

\subsection{Kodama state}

The Kodama state is a wave function in the connection representation,
given by the exponential of the CS Lagrangian,
\be
\Upsilon[A] = \exp\( -\f3{2\L}\int {\cL}_{CS}(A)\).
\ee
This state formally solves all quantum constraints
of self-dual\footnote{The restriction to the self-dual sector is crucial for the construction of the Kodama state to work.
Recently, in \cite{Magueijo:2020qcj} it was argued that the exponential of the imaginary part of the CS Lagrangian
provides an analogue of the Kodama state in the real case, i.e. it solves constraints of GR and
it continues to do so even in the modified Euler theory, which corresponds to the model considered here
in the absence of the Holst and Pontryagin terms.
The problem with this claim is that it is based on an analysis which neglects the second class constraints
of the corresponding theories.
While this is acknowledged in \cite{Magueijo:2020qcj},
it is argued there that the second class constraints are irrelevant for the issue of the Kodama state.
However, this is not true.
To see why, let us recall that the implementation of second class constraints at quantum level requires
to quantize the symplectic structure induced on their constraint surface, and not the one given a priori by the
canonical Poisson brackets.
The former can be obtained by either solving the second class constraints explicitly or
computing the Dirac brackets.
For GR one then finds that the field $\Gamma_a^i$ becomes
a differential operator $\Gamma_a^i(\hE)$ where $\hE^a_i=-\I\delta/\delta K_a^i$
and, as a result, the Kodama state manifestly fails to be a solution.
One can try to ignore this issue and quantize the theory using the naive symplectic structure,
which may work in special simple cases, like the Gupta-Bleuler procedure in quantum electrodynamics.
However, one then confronts another problem. The Hamiltonian constraint analyzed in \cite{Magueijo:2020qcj}
(which coincides with our constraint $\CH$) is {\it not} first class with respect to
the canonical Poisson brackets, or equivalently it is not the generator of any gauge symmetry.
As it should be clear from our analysis (and remains true in GR),
only a linear combination of $\CH$ with a set of second class constraints
(the one referred to as $\crH$ in our work) is first class.
(In contrast, in the symplectic structure given by Dirac brackets, $\CH$ and $\crH$ are
equivalent and are both first class.)
Thus, it is this combination that must annihilate physical states, and not the original
Hamiltonian constraint $\CH$. And it is easy to see that any version of the Kodama state is not
annihilated by such combination, so that the corresponding gauge symmetry is not implemented.
In either way, the existence of second class constraints cannot be ignored,
and ruins the construction of \cite{Magueijo:2020qcj} for both GR and the modified theory with real actions.}
Ashtekar gravity, albeit being non-normalizable \cite{Kodama:1990sc}.
It was pointed out in \cite{Alexander:2018djy} that a natural generalization of
the Kodama state taking into account the non-constant nature of $\Lambda$,
\be
\Upsilon[A,\varphi]
= \exp\( \f{3}{2\I} \int \varphi\, {\cL}_{CS}(A)\),
\label{Kodama}
\ee
where we recall that $\varphi= -\I/\L$, should play a similar role in the present modified theory of gravity.
The canonical analysis performed in the previous subsection allows us to verify this expectation and make it precise.

Choosing the representation where $\Lambda$ (or its inverse $\vp$ \eqref{defvp})
is a diagonal operator, the canonical commutation relations \eqref{cancom-sf}
are represented by taking
\be
\htP^a_i=-\I\, \frac{\delta}{\delta A_a^i},
\qquad
\hat\pi=-\I \frac{\delta}{\delta \vp}\, .
\ee
It is then easy to find that
\be
\begin{split}
\htSig^{ai} \Upsilon[A,\varphi]=&\, \(-\frac{3}{2}\, \vp\,\teps^{abc}F_{bc}^i
+\(\a+\frac32\)\tl\eps^{abc}A_{b}^i\p_c\vp\)\Upsilon[A,\varphi],
\\
\hat\pi \Upsilon[A,\varphi]=&\, -\frac{3}{2}\, {\cL}_{CS}(A)\,\Upsilon[A,\varphi].
\end{split}
\label{actSig}
\ee
The first formula in \eqref{actSig} shows that for $\alpha=-3/2$
the action of quantum operators on the Kodama state is crucially simplified,
and the result is a quantum version of the de Sitter condition \eqref{deSit}
(see also \eqref{deSit-cond}).
We interpret this fact saying that the Kodama state is \emph{localised} around the de Sitter spacetime.
It should then not come as a surprise that
the Kodama state solves the constraints only for the particular value
$\alpha=-3/2$ for which the theory admits de Sitter as solution.
For example, the action of the Gauss constraint \eqref{Gauss-sf} is given by
\be
\hat\CG^i\Upsilon[A,\varphi]=-\(\a+\frac32\)\teps^{abc}\p_a A_{b}^i\p_c\vp\, \Upsilon[A,\varphi],
\ee
where we used the Bianchi identity. Similarly, using \eqref{actSig}, it is easy to check
that the action on the Kodama state of all other
constraints\footnote{Remarkably, there are no ordering ambiguities in the corresponding
quantum operators because of the symmetries of the index contractions.}
$\hat\CV_a$, $\hat\CH$, $\hat\Pi$ and $\hat\Psi$ evaluated at $\alpha=-3/2$
vanishes as well.

This result confirms the expectation of \cite{Alexander:2018djy}
in front of the complete canonical analysis we just derived.
However, it also raises a puzzle.
The fact that the Kodama state is a simultaneous zero-eigenstate of both
$\hat\Pi$ and $\hat\Psi$ appears to be in contradiction with the results of the canonical analysis
of section~\ref{subsec-canan-sf}, where it was found that $\Pi$ and $\Psi$
form a pair of second class constraints, i.e. they are mutually non-commuting.
The resolution of this puzzle comes from the observation made above that the Kodama state
is effectively localized on the de Sitter solution,
and therefore one can argue that it is the constraints of the canonical
structure around de Sitter that should be solved by this wave function.
In this case the linearized constraints $\Pi^{(1)}$ and $\Psi^{(1)}$ do commute, see section~\ref{sec-dSSF},
so there is no tension if they vanish simultaneously.
Thus, we propose that this localization is the reason why the Kodama state is a solution in the modified theory.

\subsection{Reality conditions}

Although the self-dual theory is very nice and possesses many attractive features some of which
manifested in the previous discussion, it is not physical unless one supplements it by reality conditions
requiring the metric (and other physical fields) to be real \cite{Bengtsson:1989pe,Immirzi:1992ar}.
In the canonical language, this means that we should require the reality of
the lapse $\lapse$, of the shift $\shift^a$, and of the induced metric
\be
\Im q^{ab} \sim \Im(\delta^{ij} \tSig^a_i\tSig^b_j)=0,
\label{realmet}
\ee
whose solution can be written as
\be
\tSig^a_i=\I\(\tE^a_i-\I\eps^{ijk}\tE^a_j\chi_k\),
\label{tSigE}
\ee
which perfectly agrees with the parametrization used in the real case, see \eqref{deftSig}, \eqref{defSig} and \eqref{parameX}.
Furthermore, these conditions should be preserved by the time evolution.
This last requirement raises an issue: how should this time evolution of the reality conditions be evaluated and
what is the canonical interpretation of the resulting equation?

One can note that the reality conditions \eqref{realmet}
are identical to the (primary) simplicity constraints \eqref{covsimpl}
of the real formalism. Therefore, it is tempting to study their stability under time evolution
in the same way as for the usual constraints, namely, following the standard Dirac's formalism.
This is however problematic because they involve complex conjugate variables
which are {\it not} part of the phase space in the self-dual theory.
In particular, they have trivial Poisson brackets.
This problem occurs already in Ashtekar's self-dual GR \cite{Ashtekar:1986yd,Ashtekar:1987gu}.
There the solution was to study stability of the reality conditions using the holomorphic dynamics,
and it was recently argued in \cite{Krasnov:2020zfi} that this is in general the only reasonable way to generate their time
evolution.\footnote{In fact, in \cite{Alexandrov:2005ng} it was shown that
in the case of Ashtekar gravity the reality conditions can be implemented as second class constraints on the extended phase space.
The key for this is that for generic $\gamma$ it is possible to choose variables
such that the limit $\gamma\to \pm \I$ of the symplectic structure given by
the Dirac brackets and written in terms of these variables is smooth.
However, this relies on the knowledge of the Dirac brackets and the similarity of the canonical structures of the real and self-dual
formulations of GR, which is not the case in our theory and, more generally, in modified gravity models.
Thus, we do not expect this approach to work in the present case. We proceed instead following the general
prescription given in \cite{Krasnov:2020zfi}.}
Namely, if a reality condition has the form $\Im f=0$ where $f$ is a holomorphic function,
its time evolution is obtained as the imaginary part of the commutator with the Hamiltonian, i.e.
$\Im \{\Htot,f\}=0$.
The crucial difference with the case of the usual constraints is that the reality conditions must {\it not} be included
into the Hamiltonian generating the time evolution, which therefore remains holomorphic.
Then the resulting stability condition should be treated as usual: either it imposes a reality
condition on some Lagrange multiplier, or it induces an additional reality condition on canonical variables,
which plays a role analogous to a secondary constraint.
In the latter case, the stabilization procedure must be continued until it stops.
In the case of self-dual GR, it stops at secondary reality conditions, and one finds eventually 2 real degrees of freedom.
As we will see, the absence of the primary reality conditions
in the Hamiltonian of our modified theory leads instead to striking consequences.

Let us now apply this prescription to our model.
For simplicity we again restrict ourselves to the time gauge $\chi^i=0$. The field \eqref{tSigE} becomes purely imaginary, and
the reality conditions to be imposed on the phase space are
\be
\Re\tSig^a_i=0,
\qquad
\Re \phi=0.
\label{primrealcon}
\ee
According to the results of section \ref{subsec-canan-sf}, the time evolution is generated by the holomorphic Hamiltonian
\be
-\Htot=\CG(n)+\CD(\vec\shift)+\CH(\lapse)-\Pi\(\frac{\{\CH(\lapse),\Psi\}}{\{\Pi,\Psi\}}\).
\ee
Hence, one finds the following time evolution equations
\bea
\p_t\tSig^a_i&=&\{\tSig^a_i,\Htot\}=
\eps^{ijk}n_j\tSig^a_k+
\p_b(\shift^b\tSig^a_i)-\tSig^b_i\p_b \shift^a
\label{evolSig}\\
&&
-2\alpha\lapse \teps^{abc}\p_b\phi\(\eps^{ijk}\tSig^d_j F_{cd}^k-\frac{\I q}{\phi}\, \Sigt_c^i\)
-D_b\(\lapse\eps^{ijk}\tSig^a_j\tSig^b_k\)+\alpha \, \frac{\{\CH(\lapse),\Psi\}}{\{\Pi,\Psi\}}\, \teps^{abc}F_{bc}^i.
\nn\\
\p_t\phi &=&\{\phi,\Htot\}=\frac{\{\CH(\lapse),\Psi\}}{\{\Pi,\Psi\}}\, .
\label{evolphi}
\eea
Using these equations and taking into account the reality of the lapse and shift,
preservation of the reality conditions \eqref{primrealcon} leads to the following restrictions
\bea
&&
\eps^{ijk}(\Im n_j)\tSig^a_k
-2\alpha\lapse \teps^{abc}\p_b\phi\(\eps^{ijk}\tSig^d_j\Re F_{cd}^k-\frac{\I q}{\phi}\, \Sigt_c^i\)
-\Re \(D_b\(\lapse\eps^{ijk}\tSig^a_j\tSig^b_k\)\)
\nn\\
&&
\qquad
+\alpha \Re\( \frac{\{\CH(\lapse),\Psi\}}{\{\Pi,\Psi\}}\, \teps^{abc}F_{bc}^i\)=0,
\label{realcond1}
\\
&&
\Re \frac{\{\CH(\lapse),\Psi\}}{\{\Pi,\Psi\}}=0.
\label{realcond2}
\eea
The first equation can be split into two independent conditions.
First, after contraction with $-\hf\,\eps_{ijk}\Sigt_a^j$, we get
\be
\Im n_i=\I \tSig^a_i\p_a \lapse+ \frac{\I}{2}\, \lapse\( \nabla_a\tSig^a_i+\tSig^a_i\p_a \log q\),
\ee
where $\nabla_a\tSig^a_i=\p_a\tSig^a_i-\eps^{ijk}\Re A_a^j \,\tSig^a_k$.
This fixes the imaginary part of the Lagrange multiplier of the Gauss constraint.
Second, after contraction with $\Sigt_a^j$ and symmetrizing in $(ij)$, we get
\be
\lapse\Sigt_a^{(i}\( \eps^{j)kl} \tSig^b_k \nabla_b\tSig^a_l+2\alpha \teps^{abc}\eps^{j)kl} \Re F_{bc}^l \tSig^d_k\p_d\phi\)
= \alpha \Im\( \frac{\{\CH(\lapse),\Psi\}}{\{\Pi,\Psi\}}\) \teps^{abc}\Sigt_a^{(i}\Im F_{bc}^{j)},
\label{relcond-second}
\ee
where we took into account \eqref{realcond2}.
One can show that the commutator $\{\CH(\lapse),\Psi\}$ does not have terms with derivatives acting on the lapse.
Therefore, the lapse factorizes in this equation. Since it is non-vanishing,
the equality implies six new conditions on the complex phase space variables.
These six, together with \eqref{realcond2}, play the role of the secondary reality conditions.

It may be useful to note that the conditions \eqref{relcond-second} can be seen as an analogue of the secondary constraints
$\psi_{ij}/A$ of the real theory, see \eqref{defpsi-1},
where the role of the factor $B/A$ is played by ${\{\CH(\lapse),\Psi\}}/{\{\Pi,\Psi\}}$, i.e. the primary constraint $\phi$
is replaced by the secondary constraint $\Psi$. This replacement is a consequence of the fact that
the dynamics is generated by the holomorphic Hamiltonian.

So far everything looks fine and in line with the analysis of the real case.
However, while in the real theory the secondary constraints are stabilized fixing Lagrange multipliers
(of the primary simplicity constraints) and without introducing tertiary constraints,
the situation is dramatically different in the self-dual theory.
The point is that there are no Lagrange multipliers left to be fixed: there are no simplicity constraints here,
and all the multipliers still free correspond to the first class
constraints which are known to be generators of gauge transformations.
Hence, the evolution of the secondary reality conditions gives rise to equations
which either hold automatically, which is practically impossible given the complicated form
of \eqref{realcond2} and \eqref{relcond-second}, or lead to new restrictions on the phase space variables.
As a result, the stabilization procedure continues and never stops,
as in the case of modified gravity models considered in \cite{Krasnov:2020zfi}.
Thus, we are forced to conclude that the imposition of the reality conditions on the self-dual theory
makes it inconsistent killing all degrees of freedom.

\section{Conclusions}
\label{sec-concl}

In this paper we have provided a complete canonical analysis of the minimal varying $\Lambda$ theories of
\cite{Alexander:2018djy,Alexander:2019ctv,Alexander:2019wne,Magueijo:2019vmk}, with general couplings
that include the Einstein-Cartan-Holst term plus the
Euler and Pontryagin topological terms weighted by $1/\Lambda$, which is considered as a dynamical scalar field.
We have shown that for the coupling constants satisfying the relation \eqref{condparam},
the model has three degrees of freedom and if not, their number is increased to five.
In both cases we have found explicit expressions for all constraints
and identified whether they are first or second class.
The canonical structure however changes when one considers these theories linearized around the de Sitter background.
Then a new gauge symmetry emerges which reduces the number of degrees of freedom by one.

We have also identified the canonical structure of the self-dual theory described by a holomorphic action.
In this case the analysis gives two complex degrees of freedom. This result remains true on the de Sitter background
despite the emergence of the same gauge symmetry as in the real case.
We have verified that a natural generalization of the Kodama state is annihilated by all quantum constraint operators.
On the other hand, an attempt to implement reality conditions leads to inconsistencies with the holomorphic dynamics,
which makes problematic the use of the self-dual theory for physical applications.

Let us conclude with perspectives on future work.
First, the action of our model could be further generalized including a coupling to the Nieh-Yan term,
the last remaining of the six possible terms that can be written in differential forms.
We have shown that including this term with the same $1/\L$ coupling rules out the de Sitter solution,
and explained that it can be restored if one accepts a logarithmic coupling instead.
In either case, the methods and strategy presented in this paper can be applied to analyse
the canonical structure and resulting number of degrees of freedom.

Second, we did not perform the analysis in the case when the spatial gradient of $\L$
vanishes but the parameters are not restricted by $\ab=0$.
We leave the study of this sector
as an open question.

We also have not attempted to identify the physical interpretation of
the additional degrees of freedom.
Because of the presence of a metric field and diffeomorphism invariance, we expect two degrees of freedom
to be always represented by the massless spin-2 graviton mode, with the caveat that
since the Minkowski spacetime is not a solution for this theory,
the transverse-traceless decomposition is more complicated.
For the special case $\ab=0$, it is natural to expect that the unique extra degree of freedom
should be described by a scalar whose gradient gives one of the two spin-1 parts of torsion.
We thus anticipated that
this case may correspond to
a scalar-tensor theory.
For the general case with three extra degrees of freedom, there are multiple options.
What seems unlikely is to have three scalars, because there is no scalar mode in the spin-2 part of torsion.
It may be a scalar and a massless spin-1 or spin-2 mode, but this remains to be seen.

Finally, it is worth mentioning that the phenomenological implications and viability of these models is under investigation.
Existing results regarding the behaviour of a subset of the models considered
in this paper on cosmological backgrounds describing expansion and propagation of gravitational waves
suggest that there exist regions of the parameter space which are consistent
with experimental constraints \cite{Alexander:2020umk}. Looking towards future work,
it is conceivable that deviations from general relativity due to the spacetime variation
of $\Lambda$ could lead to observable effects in precision tests of gravity such
as gravitational wave astronomy \cite{Barrientos:2019awg} and
post-Newtonian gravitational effects in the solar system or compact objects such as neutron stars.

\section*{Acknowledgements}

T.Z. is supported by the Grant Agency of the Czech Republic GA\v{C}R grant 20-28525S.
T.Z thanks Jo\~{a}o Magueijo and Pavel Jirou\v{s}ek for helpful discussions.

\appendix

\section{Torsion and de Sitter}
\label{ap-deSitter}

In this appendix we rewrite the equations determining the de Sitter solution in terms of the tetrad and $\Lambda$,
eliminating $\om^{IJ}$ through the solution \eqref{deSit-tor} for the torsion.
We follow the conventions of \cite{Speziale:2018cvy} and refer the reader to this paper and to the classical
reference \cite{Hehl:1976kj} for more details.

First, we recall the definition of the contorsion tensor $C^{IJ}$,
which describes the deviation from the torsion-free Levi-Civita connection $\om^{IJ}(e)$,
\be
\om^{IJ}=\om^{IJ}(e)+C^{IJ}.
\ee
It is related to the torsion $T^I=D^{(\om)} e^I$ by
\be
T^I=C^{IJ}\w e_J,
\qquad T^\r{}_{\m\n}=e^\r_I \, T^I{}_{\m\n}= 2C_{[\m,}{}^\r{}_{\n]}.
\ee
The advantage of using contorsion is that it allows us to write the decomposition of the curvature in a simple way
\be
F^{IJ}(\om) = F^{IJ}(\omega(e)) + D^{\om(e)}C^{IJ} + C^{IK}\w C_K{}^J,
\ee
or explicitly in components,
\be
\label{RiemC}
R_{\m\n\r\s} = R_{\m\n\r \s}(g)+ 2\og\nabla_{[\r}C_{\s],\m\n}  + 2C_{[\r,\m\l}C_{\s],}{}^\l{}_{\n},
\ee
where $\og\nabla_\mu$ denotes the torsion free covariant derivative.

The solution for torsion  \eqref{deSit-tor} corresponding to the de Sitter ansatz \eqref{deSit} gives for the contorsion
\be
C_{\m,\n\r} = \f12\, g_{\m[\n}\p_{\r]}\log\L.
\ee
Plugging this solution into \eqref{RiemC}, one finds
\be
\begin{split}
R_{\m\n\r\s} =&\, R_{\m\n\r\s}(g) +g_{\m[\s}\ov{\na}{g}_{\r]}\p_\n\log\L - g_{\n[\s}\ov{\na}{g}_{\r]}\p_\m\log\L
\\
&\, + \f12\left( g_{\m[\r} \p_{\s]}\log\L \, \p_{\n}\log\L - g_{\n[\r} \p_{\s]}\log\L \, \p_{\m}\log\L
-  g_{\m[\r}g_{\s]\n} (\p\log\L)^2  \right).
\end{split}
\ee
Equating this expression to the original ansatz \eqref{deSit} gives $C_{\m\n\r\s}(g)=0$, and
\be
{R}_{\mu\nu}(g) - \ov{\na}{g}_{\nu}\p_\mu\log\L -\hf\, g_{\mu\nu} \ov{\square}{g} \log\L
+ \hf\(\p_\mu\log\L \,\p_\nu\log\L -g_{\mu\nu}(\p\log\L)^2\) = \L g_{\mu\nu}.
\label{Ricci-eq}
\ee
The left-hand side can be recognized as a result of the conformal transformation $g_{\mu\nu}\mapsto \L g_{\mu\nu}$,
and using the conformal invariance of $C^\m{}_{\n\r\s}$, one arrives at the equations \eqref{Weyl-eq} presented in the main text.

\section{Useful commutation relations}
\label{ap-useful}

In this appendix we collect various commutation relations that appear in the calculations of section \ref{sec-canan}.
First, the canonical commutation relations are given by
\be
\begin{split}
\{ K_a^i(\xc),\tP^b_j(\yc)\}=&\, \delta_a^b\delta^i_j\,\delta^{(3)}(\xc,\yc),
\\
\{ \Gamma_a^i(\xc),\tQ^b_j(\yc)\}=&\, \delta_a^b\delta^i_j\,\delta^{(3)}(\xc,\yc),
\\
\{ \Lambda(\xc),\pi(\yc)\}=&\,\delta^{(3)}(\xc,\yc)
\end{split}
\label{canPB}
\ee
with all other Poisson brackets between canonical variables vanishing.

The Poisson brackets of the constraints generating local rotations and boosts with the curvature tensor are given by
\bea
\{ \CR(n), F_{ab}^{0i}\}&=& {\eps^i}_{jk}n^j F_{ab}^{0k},
\qquad
\{ \CR(n), F_{ab}^{i}\}= {\eps^i}_{jk}n^j F_{ab}^{k},
\\
\{ \CK(n), F_{ab}^{0i}\}&=& {\eps^i}_{jk}n^j F_{ab}^{k},
\qquad
\{ \CK(n), F_{ab}^{i}\}= -{\eps^i}_{jk}n^j F_{ab}^{0k},
\eea
where the components of the curvature have the following explicit expressions in terms of the canonical variables
\be
\begin{split}
F_{ab}^{0i}=&\, \p_a K_b^i-\p_b K_a^i -{\eps^i}_{jk}\(K_{a}^j\Gamma_{b}^k+\Gamma_{a}^j K_{b}^k\),
\\
F_{ab}^{i}=&\, \p_a\Gamma_b^i-\p_b\Gamma_a^i +{\eps^i}_{jk}\(K_{a}^j K_{b}^k-\Gamma_{a}^j\Gamma_{b}^k\) .
\end{split}
\label{compF}
\ee
Using these commutation relations, we immediately obtain
\bea
\{ \CR(n), \tcP^a_i\}&=& {\eps_{ij}}^k n^j \tcP^a_k,
\qquad\ \
\{ \CR(n),\tcQ^a_i\}= {\eps_{ij}}^k n^j \tcQ^a_k,
\\
\{ \CK(n), \tcP^a_i\}&=& -{\eps_{ij}}^k n^j \tcQ^a_k,
\qquad
\{ \CK(n), \tcQ^a_i\}= {\eps_{ij}}^k n^j \tcP^a_k,
\eea
where $\tcP^a_i$ and $\tcQ^a_i$ are defined in \eqref{momenta}, whereas for the densitized triad \eqref{defE2}
one finds
\be
\{ \CR(n), \tE^a_i\}= {\eps_{ij}}^k n^j \tE^a_k,
\qquad
\{\CK(n),\tE^a_i\}=-\frac{{\eps_{ij}}^k n^j \phi^a_k}{1+1/\gamma^2}\,.
\ee

The densitized triad has also the following Poisson brackets with the primary constraints \eqref{primcon} and with itself
\begin{subequations}
\bea
\{\phi(\lambda),\tE^a_i\}&=&\frac{2\abb}{1+1/\gamma^2}\,\teps^{abc}\lambda_b^i\p_c\Lambda^{-1},
\label{com-phiE}
\\
\{\tE(\vec f),\tE(\vec g)\}&=&\frac{2\ab}{(1+1/\gamma^2)^2}\int\de^3 \xc\,\teps^{abc} f_a^ig_b^i\,\p_c\Lambda^{-1},
\label{com-EE}
\eea
\label{commE}
\end{subequations}
where the parameters $\abb$ and $\ab$ are given by
\begin{subequations}
\bea
\abb&=&\alpha\(1-\frac{1}{\gamma^2}+\frac{2}{\gamma\gamma'}\)=\(1-\frac{1}{\gamma^2}\)\aE+\frac{2\aP}{\gamma},
\label{defabb}
\\
\ab&=&\frac{\alpha}{\gamma'}\(1-\frac{1}{\gamma^2}-\frac{2\gamma'}{\gamma}\)=\(1-\frac{1}{\gamma^2}\)\aP-\frac{2\aE}{\gamma}.
\label{defa0}
\eea
\label{twoparam}
\end{subequations}
It is straightforward to check that the vector $v_i$ \eqref{defv}, playing an important role in our analysis,
transforms as a scalar density under spatial diffeomorphisms, i.e.
\be
\{\CD(\vec\shift),v_i\}=\p_a(\shift^a v_i).
\label{com-Dv}
\ee

Finally, we present the commutation relations of the quantities $A_{ij}$ and $B_{ij}$ \eqref{defAB}, which
determine the secondary constraints of the model, with all primary constraints except the Hamiltonian one $\CH$
since the latter is not necessary for understanding the canonical structure.
The corresponding Poisson brackets are found to be
\begin{subequations}
\bea
\{ \CD(\vec\shift), A_{ij}\} &=& \shift^a\p_a A_{ij},
\label{com-DA}
\\
\{ \CR(n), A_{ij}\} &=& 2{\eps_{(ik}}^l n^k A_{lj)},
\label{com-RA}
\\
\{ \CK(m), A_{ij}\} &\approx & \Lambda^{-1} \Et_{a(i}{\eps_{j)}}^{kl}m_k\(\tcP^a_l+\tfrac{1}{\gamma}\,\tcQ^a_{(i}\),
\label{com-KA}
\\
\{ \phi(\vec\lambda), A_{ij}\}
&=&
\frac{2}{q\Lambda^{2}}\,{\eps_{(i}}^{mn}\tE^a_m
\(\abb\, {\eps_{j)k}}^lK_a^k\lambda_{nl}-\ab\(\p_a\lambda_{nj)}-{\eps_{j)k}}^l\Gamma_a^k\lambda_{nl}\)\)
-\frac{2\ab}{\Lambda^{2}}\,\teps^{abc}\Et_{a(i}\lambda_{j)k}\p_b\Et_c^k
\nn\\
&&
-\frac{2\abb}{q\Lambda(1+1/\gamma^2)}\,{\eps_{(i}}^{kl}\lambda_{km}\tE^a_l\Et_b^m\(\tcQ^b_{j)}-\tfrac{1}{\gamma}\,\tcP^b_{j)}\)\p_a\Lambda^{-1},
\label{com-phiA}
\eea
\label{commA}
\end{subequations}
\begin{subequations}
\bea
\{ \CD(\vec\shift), B_{ij}\} &=& \p_a (\shift^a B_{ij}),
\label{com-DB}
\\
\{ \CR(n), B_{ij}\} &=& 2{\eps_{(ik}}^l n^k B_{lj)},
\label{com-RB}
\\
\{ \CK(m), B_{ij}\} &\approx & -\(1+\tfrac{1}{\gamma^2}\)\(\delta_{(ik}\tE^a_{j)}-\delta_{ij}\tE^a_k\){\eps^k}_{mn} m^m K_a^n
\nn\\
&&-\frac{\abb \p_a\Lambda^{-1}}{1+1/\gamma^2}\,
\teps^{bcd}\Et_b^{(i}\tE^a_k\(m^{j)} \(F^{0k}_{cd}+\tfrac{1}{\gamma}\, F_{cd}^{k}\)
-m^k\(F^{0j}_{cd}+\tfrac{1}{\gamma}\, F_{cd}^{j}\) \),
\label{com-KB}
\\
\{ \phi(\vec\lambda), B_{ij}\}
&=&-\(1+\tfrac{1}{\gamma^2}\)\(\lambda_{ij}-\delta_{ij}{\lambda_k}^k\)
\nn\\
&&
+2\abb \Bigl[\teps^{bcd}{\eps_{(i}}^{kl}\lambda_{lm}\Bigl(\Et_{bj)}\(\tE^a_k\Et_d^m \p_a\p_c\Lambda^{-1}-\Et_a^m\nabla_d \tE^a_k\p_c\Lambda^{-1}\)
-\Et_{aj)}\Et_d^m\nabla_b\tE^a_k\p_c\Lambda^{-1}\Bigr)
\nn\\
&&
+q^{-1}{\eps_{(i}}^{mn}\lambda_{mp}\Et_c^p{\eps_{j)}}^{kl}\tE^a_n\tE^b_k\nabla_b\tE^c_l\p_a\Lambda^{-1}
\Bigr]
\nn\\
&&
+\frac{2\abb^2\p_a\Lambda^{-1}\p_b\Lambda^{-1}}{q\(1+1/\gamma^2\)^2}\,
\eps^{(ikl}\eps^{j)mn}\tE^a_k\tE^b_m\teps^{cdf}\(F^{l}_{cd}-\tfrac{1}{\gamma}\, F_{cd}^{0l}\)
\Et_f^p\lambda_{np}.
\label{com-phiB}
\eea
\label{commB}
\end{subequations}

\section{Constraint algebra}
\label{ap-conalg}

\subsection{General case}
\label{ap-conalg1}

The algebra of primary constraints can be computed in the straightforward way using
the commutation relations presented in the previous appendix.
The commutators which are weakly vanishing are the following
\begin{subequations}
\bea
\{\CR(n), \pi(\rho)\} &=& \{\CK(m), \pi(\rho)\}=0,
\\
\{ \CR(n), \CR(m)\} &=& -\CR(n\times m),
\\
\{ \CR(n), \CK(m)\} &=& -\CK(n\times m),
\\
\{ \CK(n), \CK(m)\} &=& \CR(n\times m),
\\
\{ \CR(n), \phi(\vec \lambda)\} &=& -\phi(n\times \vec \lambda),
\\
\{ \CR(n), \CH(\lapse)\} &=&0,
\\
\{ \CK(m), \CH(\lapse)\} &=&  \phi(\lapse m\times \vec f)+\CV(\lapse m^i\tE_i),
\label{com-cKH}
\\
\{ \CD(\vec\shift), \CR(n)\} &=& -\CR(\shift^a\p_a n),
\\
\{ \CD(\vec\shift), \CK(m)\} &=& -\CK(\shift^a\p_a m),
\\
\{ \CD(\vec\shift), D(\vec M)\} &=& -\CD([\vec\shift,\vec M]),
\\
\{ \CD(\vec\shift), \CH(\lapse)\} &=& -\CH(\CL_{\vec\shift}\lapse),
\\
\{ \CD(\vec\shift), \pi(\rho)\} &=& -\pi(\shift^a\p_a\rho),
\\
\{ \CD(\vec\shift), \phi(\vec \lambda)\} &=& -\phi(\CL_{\vec\shift} \vec\lambda),
\\
\{\CH(\lapse),\CH(\Mt)\}&=& 0,
\label{com-HH}
\eea
\label{1cl-alg}
\end{subequations}
where
\bea
(n\times m)^i&=&{\eps^i}_{jk} n^j m^k,
\\
f^a_i &=& \tfrac{1}{1+1/\gamma^2}\,\({\eps^{ij}}_k \tE^b_j  \(F^{k}_{ab}-\tfrac{1}{\gamma}\, F_{ab}^{0k}\)-\Lambda q\Et_a^i\),
\\
{[\vec\shift,\vec M]}{}^a &=&\shift^b\p_b M^a-M^b\p_b \shift^a,
\\
\CL_{\vec\shift}\lapse &=& \shift^b\p_b \lapse-\lapse\p_b \shift^b,
\\
(\CL_{\vec\shift} \vec\lambda)_a^i &=&\shift^b\p_b \lambda_a^i+\lambda_b^i\p_a \shift^b.
\eea
The non-vanishing commutators are given by
\begin{subequations}
\bea
\{ \CK(m), \phi(\vec \lambda)\} &=&  -\(1+\tfrac{1}{\gamma^2}\)\int\de\xc^3\, {\eps_{ij}}^k m^i \lambda_a^j \tE^a_k,
\label{com-Kphi}
\\
\{\phi(\vec \lambda),\phi(\vec \mu)\} &=& -2\ab\int\de^3 \xc\,\teps^{abc} \lambda_a^i\mu_b^i\,\p_c\Lambda^{-1},
\label{com-phiphi}
\\
\{ \pi(\rho), \phi(\vec \lambda)\} &=& -\int\de\xc^3\, \rho\lambda_a^i \, \Lambda^{-1} \(\tcQ^a_i-\tfrac{1}{\gamma}\,\tcP^a_i\),
\label{com-piphi}
\\
\{\phi(\vec \lambda),\CH(\lapse)\}&=& -\int\de\xc^3\,\lapse\biggl[\(1+\tfrac{1}{\gamma^2}\)q\,\teps^{abc}\Et_c^i
\nabla_a\lambda_b^i
\nn\\
&&\qquad
+\frac{2\abb\p_a\Lambda^{-1}}{1+1/\gamma^2}\,\teps^{abc} \lambda_b^i
\(\tE^d_j \(F^{k}_{cd}-\tfrac{1}{\gamma}\, F_{cd}^{0k}\)-\Lambda q \Et_c^i\)\biggr],
\label{com-phiH}
\\
\{ \pi(\rho), \CH(\lapse)\} &=&\int\de\xc^3\rho\lapse\, \CE,
\label{com-piH}
\eea
\label{2cl-alg}
\end{subequations}
where $\nabla_a$ is the covariant derivative defined by the $SU(2)$ connection $\Gamma_a^i$
\be
\nabla_a V^i=\p_a V^i-{\eps^i}_{jk}\Gamma_a^j V^k,
\label{defnabla}
\ee
and we introduced the function
\be
\CE= q+\frac{q(\tcP^a_i+\frac{1}{\gamma}\,\tcQ^a_i) \Et_a^i}{1+1/\gamma^2}
-\frac{{\eps^{ij}}_k(\tcP^a_i+\frac{1}{\gamma}\,\tcQ^a_i)\tE^b_j}{(1+1/\gamma^2)\Lambda} \(F^{k}_{ab}-\tfrac{1}{\gamma}\, F_{ab}^{0k}\).
\label{defCE}
\ee

For our analysis, it is convenient to split the commutation relations \eqref{com-piphi} and \eqref{com-phiH} into
two contributions corresponding to symmetric and antisymmetric parts of $\lambda^{ij}=\lambda^{a,i}\tE_a^j$. One then obtains
\begin{subequations}
\bea
\{ \pi(\rho), \phi(\vec \lambda)\} &=& -\int\de\xc^3 \(\lambda^{(ij)}A_{ij}-\(1+\tfrac{1}{\gamma^2}\){\eps_{ij}}^k\lambda^{[ij]}\CA_k\)\rho ,
\label{com-piphi-split}
\\
\{\phi(\vec \lambda),\CH(\lapse)\}&=& -\int\de\xc^3\(\lambda^{(ij)}B_{ij}+\(1+\tfrac{1}{\gamma^2}\){\eps_{ij}}^k\lambda^{[ij]}\CB^+_k\)\lapse ,
\label{com-phiH-split}
\eea
\label{2cl-alg-split}
\end{subequations}
where
\begin{subequations}
\bea
A_{ij}&=& \Lambda^{-1}\Et_{a(i}\(\tcQ^a_{j)}-\tfrac{1}{\gamma}\,\tcP^a_{j)}\),
\\
B_{ij}&=& -\(1+\tfrac{1}{\gamma^2}\)\Et_{b(i}{\eps_{j)}}^{kl}\tE^a_k\nabla_a\tE^b_l
+\frac{\abb \p_a\Lambda^{-1}}{1+1/\gamma^2}\, \teps^{bcd}\Et_b^{(i}{\eps^{j)k}}_l\tE^a_k\(F^{l}_{cd}-\tfrac{1}{\gamma}\, F_{cd}^{0l}\) ,
\eea
\label{defAB}
\end{subequations}
and
\begin{subequations}
\begin{align}
\CA_i&= \frac{{\eps_{ij}}^k \Et_a^j(\tcQ^a_k-\frac{1}{\gamma}\,\tcP^a_k)}{2\Lambda(1+1/\gamma^2)}\,,
\label{defCAnew}
\\
\CB^\pm_i&=\pm\tE^a_i \p_a+\frac{1}{2q}\, \p_a(q\tE^a_i)
-\hf\,\eps_{ijk}\Gamma_a^j\tE^a_k
+\frac{\abb\p_a\Lambda^{-1}}{q(1+1/\gamma^2)^2}\, {\eps^{jk}}_l\tE^a_j\tE^b_i\tE^c_k \(F^{l}_{bc}-\tfrac{1}{\gamma}\, F_{bc}^{0l}\).
\label{defCBnew}
\end{align}
\label{defCAB}
\end{subequations}
Note that $\CB^\pm_i$ are differential operators. Either of them can be used to express the commutator \eqref{com-phiH-split},
the two expressions being related by integration by parts.

\subsection{Self-dual case}
\label{ap-conalg2}

In the self dual case, the constraint algebra is given by
\begin{subequations}
\bea
\{\Pi(\rho),\Pi(\vr)\}&=& 0,
\\
\{\CG(n),\Pi(\rho)\}&=&0,
\\
\{ \CG(n), \CG(m)\} &=& -\CG(n\times m),
\\
\{ \CG(n), \CH(\lapse)\} &=& 0.
\\
\{ \CD(\vec\shift), \CG(n)\} &=& -\CG(N^a\p_a n),
\\
\{ \CD(\vec\shift), \CD(\vec M)\} &=& -\CD([\vec\shift,\vec M]),
\\
\{ \CD(\vec\shift), \CH(\lapse)\} &=& -\CH(\CL_{\vec\shift}\lapse),
\\
\{ \CD(\vec\shift), \Pi(\rho)\} &=& -\pi(N^a\p_a\rho),
\\
\{\CH(\lapse),\CH(\Mt)\}&=& -\CV(\vec K_{N,M}),
\label{defHH-SD}
\eea
\end{subequations}
where
\be
K_{N,M}^a
=-\tSig^a_i\tSig^b_i (\Nt\p_b\Mt-\Mt\p_b\Nt),
\label{defKMN}
\ee
and the only commutator which is not weakly vanishing is the following
\be
\{\Pi(\rho),\CH(\lapse)\}= -2\alpha \CH(\lapse\rho\vp^{-1})
+\int \de^3 \xc\, \lapse\[\I(1+2\alpha)\,\frac{q}{\vp^2}+\alpha\teps^{abc}\eps_{ijk} F_{ab}^i \tSig^d_j F_{cd}^k\].
\label{com-PiH}
\ee

\providecommand{\href}[2]{#2}\begingroup\raggedright\endgroup


\end{document}